\documentclass[aps,prl,twocolumn,floatfix,superscriptaddress,showpacs,amsmath,amssymb]{revtex4}
\usepackage{graphicx}
\usepackage{epsfig}
\allowdisplaybreaks

\newcommand{\be}{\begin{equation}}
\newcommand{\ee}{\end{equation}}
\newcommand{\bea}{\begin{eqnarray}}
\newcommand{\eea}{\end{eqnarray}}
\newcommand{\ba}{\begin{eqnarray*}}
\newcommand{\ea}{\end{eqnarray*}}

\newcommand{\m}[1]{\mathcal{#1}}

\begin{document}

\title{Transient Orthogonality Catastrophe in a Time Dependent Nonequilibrium Environment}
\author{Marco Schir\'o}
\affiliation{Princeton
  Center for Theoretical Science and Department of Physics, Joseph Henry
  Laboratories, Princeton University, Princeton, NJ
  08544}
\affiliation{Department of Chemistry, Columbia University, New York, New York 10027, U.S.A.}
\affiliation{Institut de Physique Th\'{e}orique, CEA, CNRS-URA 2306, F-91191, 
Gif-sur-Yvette, France}  
\author{Aditi Mitra}
\affiliation{Department of Physics, New York University, 4 Washington
  Place, New York, New York 10003, USA}

\date{\today}

\pacs{72.10.Pm,05.70.Ln,72.15.Qm}

\begin{abstract}
We study the response of a highly-excited time dependent quantum many-body state to a sudden local perturbation, a sort of orthogonality catastrophe problem in a transient non-equilibrium environment. To this extent we consider, as key quantity, the overlap between time dependent wave-functions, that we write in terms of a novel two-time correlator generalizing the standard Loschmidt Echo. We discuss its physical meaning, general properties, and its connection with experimentally measurable quantities probed through non-equilibrium Ramsey interferometry schemes. Then we present explicit calculations for a one dimensional interacting Fermi system brought out of equilibrium by a sudden change of the interaction, and perturbed by the switching on of a local static potential.
We show that different scattering processes give rise to remarkably different behaviors at long times, quite opposite from the equilibrium situation. In particular, while the forward scattering contribution retains its power law structure even in the presence of a large non-equilibrium perturbation, with an exponent that is strongly affected by the transient nature of the bath, the backscattering term is a source of non-linearity which generates an exponential decay in time of the Loschmidt Echo, reminiscent of an effective thermal behavior.
\end{abstract}

\maketitle
\textit{Introduction - }The response of gapless quantum many-body systems to sudden local perturbations is a remarkably non-linear phenomenon, even a weak disturbance substantially changes the structure of the many-body state. Signatures of this \emph{orthogonality catastrophe} (OC) emerge in various condensed matter settings~\cite{Anderson_prl67}, from X-ray spectra in metals~\cite{XrayEdgeND} and Luttinger Liquids (LL)~\cite{GogolinPRL93,KaneFisherPRL92,KaneFisherPRB92,MedenEtAlPRB98} to the physics of the Kondo Effect~\cite{AY,AY2,Tureci_prl11,Imamoglu_nature11,GoldsteinPRB12} and typically results in power-law decays of dynamical correlations. Recently, impressive experimental developments with ultracold atomic gases~\cite{Bloch_rmp08}, have made possible to create and probe local excitations in a quantum many-body system with single-site and real-time resolution~\cite{Weitenberg11,FukuharaNatPhys13}, bringing fresh new input to this venerable problem~\cite{KnapetalPRX12}.
While most of the attention has been traditionally devoted to perturbations acting on systems in their ground state or, more recently, in driven stationary non-equilibrium conditions~\cite{NgPRB96,MuzykantskiiEtAlPRL03,BrauneckerPRB03,AbaninLevitovPRL04,MitraMillisPRB07,SegalPRB07,DallaTorreetalPRB2012},  much less is known about the response of explicitly time dependent quantum states, such as, for example, those obtained by rapidly changing in time some parameter of an otherwise isolated system.
The problem is of current experimental relevance since ultracold gases have proven to be natural laboratories where dynamical quantum correlations can be probed in the time domain. In addition, it also raises a number of intriguing theoretical questions. A coherent time dependent excitation, such as a sudden global quench, creates an effective non-equilibrium time-dependent bath for the local degrees of freedom. What is the effect of such an environment on the OC phenomenon and its associated power laws? For a generic, non-integrable, quantum many-body system one might expect this environment to be, at sufficiently long times, effectively thermal,
turning the power law decay of the OC correlator into an exponential. Yet, strongly interacting quantum systems may often get trapped into long-lived metastable prethermal states 
which may still show genuine quantum correlations~\cite{Berges04,Kehrein_prl08,GringScience12,Karrasch_PRL12,MitraPRB13}.
Investigating the local spectral properties of these transient states of non-equilibrium quantum matter is among the purposes of this Letter.

\textit{Transient OC Protocol -} 
We begin with a general discussion of the non-equilibrium protocol that will be the focus of this paper. We consider a quantum many-body system initially prepared at time $t_0=0$ in the ground state $\vert\Psi_0\rangle$ of some Hamiltonian $\m{H}_0$. We then let the system evolve up to some time $t>t_0$ with a new Hamiltonian $\m{H}$,  i.e. $\vert\Psi(t)\rangle=
e^{-i\m{H}t}\vert\Psi_0\rangle$, which differs from $\m{H}_0$ by a sudden change of some global parameter.
This global quantum quench injects extensive energy into the system and triggers transient non-equilibrium dynamics.
To gain insights into the structure of this transient state we will then switch-on a local perturbation $V_{\rm loc}$ for an interval $\tau$ between $t$ and $t'=t+\tau$ and compare the states obtained at time $t'$, respectively in presence or absence of the local perturbation. In the spirit of an OC problem we study the overlap
\bea\label{eqn:Ddef}
\m{D}(t',t)\equiv \langle\Psi(t')\vert\Psi_{t+}(t')\rangle
\eea
where $\vert\Psi_{t+}(t')\rangle\equiv e^{-i\,\m{H}_+(t'-t)}\vert\Psi(t)\rangle$, with $\m{H}_+=\m{H}+V_{\rm loc}$.
This overlap can be written suggestively as a two-time dynamical correlator
\bea\label{eqn:Ddef2}
\m{D}(t',t)=
 \langle\Psi(t)\vert\,e^{i\,\m{H}(t'-t)}\,e^{-i\,\m{H}_+(t'-t)}\vert\Psi(t)\rangle
\eea
One immediately sees that when the initial state $\vert\Psi_0\rangle$ is the ground state of $\m{H}$, then $\m{D}(t',t)\equiv D_{\rm eq}(\tau)=\langle\Psi_0\vert e^{i\,\m{H}\tau}\,e^{-i\,\m{H}_+\tau}\vert\Psi_0\rangle$, i.e. it becomes time-translational invariant and reduces to the familiar OC correlator, also known as core-hole Green's function in the X-ray edge problem~\cite{XrayEdgeND} or Loschmidt Echo amplitude~\cite{Peres_PRA84,Gorin_PhysRep06,HeylKehreinPRB12,DoraetalPRL2013}, which recently has attracted renewed interest in the context of work statistics~\cite{Silva_work_statistics,GambassiSilvaPRL12,HeylKehreinPRL12,
HeylKehreinPolkovnikovPRL13,Vasseur_etalPRL13}. In equilibrium, the large time asymptotics of $D_{\rm eq}(\tau)$ gives rich information
on the ground state $\vert\Psi_0\rangle$ and its low-lying excitations. In particular a power-law decay reflects an orthogonality catastrophe in the low-energy sector induced by the local perturbation, mirroring the one introduced by Anderson for stationary (ground) states~\cite{Anderson_prl67}. Considering the enormous principle importance of this phenomenon for equilibrium quantum many-body physics, it is natural to investigate its fate for time dependent excited states, as we are going to do in the following.
Before turning to an example, it is useful to discuss some general features of our transient OC correlator. We start by writing it in terms of the exact eigenstates of $\m{H}$, $\m{H}\vert\Phi_n\rangle=E_{n}\vert\Phi_n\rangle$~\cite{SM_ImpurityLL}
\be\label{eqn:D_exact}
D(\tau;t)=\sum_{nm}\rho_{nm}(t)\,\langle\Phi_m\vert
e^{i\,\m{H}\tau}\,e^{-i\,\m{H}_+\tau}\vert\Phi_n\rangle
\ee
where $\rho_{nm}(t)=\langle\Phi_n\vert\Psi(t)\rangle\langle\Psi(t)\vert\Phi_m\rangle$. Differently from the equilibrium case, here both diagonal and off-diagonal matrix elements contribute to $D(\tau;t)$ with a time-dependent amplitude $\rho_{nm}(t)$ encoding information about the state of the system before the switching on of the local perturbation. The result highlights the very nature of our transient correlator as a sensitive probe of OC in excited many-body states.  By averaging Eq.~(\ref{eqn:D_exact}) over the waiting time $t$ and taking the Fourier transform with respect to $\tau$ we obtain~\cite{SM_ImpurityLL}
$P(W)\equiv\int\,d\tau\,e^{iW\tau}\,\overline{D(\tau ;t)}$ as
\bea\label{eqn:P_w}
P(W)=\sum_{n\alpha}\,\delta(W-\tilde{E}_\alpha+E_n)\,
\vert\langle\Phi_n\vert\tilde{\Phi}_{\alpha}\rangle\vert^2\,\rho_{nn}(0)
\eea
where we have introduced the eigenstates of the full Hamiltonian, $\m{H}_{+}\vert\tilde{\Phi}_{\alpha}\rangle=\tilde{E}_{\alpha}\vert\tilde{\Phi}_{\alpha}\rangle$, in the presence of the local perturbation. $P(W)$ is the probability distribution of the work done, starting 
from a \emph{non-equilibrium state} described by the diagonal ensemble
$\rho_{nn}$, and suddenly switching on the local potential. This extends to the non-equilibrium case the connection 
between the OC correlator and work statistics and represents an interesting result on its own, in view of recent theoretical 
interest in characterizing the work statistics and its universal properties in non-thermal 
ensembles~\cite{HickeyGenway_arxiv14,PalmaiSotiriadis_arxiv14}.
While it is known that in equilibrium the work statistics of local quenches shows zero-temperature edge 
singularities\cite{Silva_work_statistics,Tureci_prl11}, we will see below how this is modified in the presence of bulk excitations.

Finally, it is interesting to discuss experimental protocols to measure the transient OC correlator. Recent proposals outline how the equilibrium OC correlator may be measured~\cite{MicheliEtAlPRL04,GooldEtAlPRA11,KnapetalPRX12,SindonaEtAlPRL13,KnapEtAlPRL13,DoraetalPRL2013,PaternostroPRL13,FazioPRL13}.
The key is to use an auxiliary two-level system (TLS) or qubit, coupled to the system through the local perturbation, $\m{H}[\sigma^z]=\m{H}+V_{\rm loc}\left(1+\sigma^z\right)/2$. Cold atoms and other hybrid systems, such as circuit QED units, represent the natural platforms to realize this. Here we extend these ideas to design non-equilibrium Ramsey interferometry schemes to manipulate the TLS in such a way as to obtain $D(t';t)$ out of simple local TLS measurements. For example, the real part of the transient OC correlator can be obtained by measuring the $x-$component of TLS magnetization~\cite{SM_ImpurityLL},
\be
\mbox{Re}D(t';t)=\langle\Phi_{t}(t')\vert\,\sigma^x\vert\Phi_{t}(t')\rangle
\ee
where $\vert\Phi_{t}(t')\rangle$ is a state obtained with a specific protocol involving (i) a first evolution up to time $t$ (ii) a Ramsey $\pi/2$ pulse and (iii) a second time evolution up to time $t'=t+\tau$. Alternatively, one can obtain $D(t';t)$ from a TLS dynamical correlator, a non-equilibrium analog of the core-hole Green's function in the X-ray edge problem~\cite{SM_ImpurityLL}.

\textit{Quenched Luttinger Liquid in a Transient Local Potential- } We now focus our attention on a 1D spinless interacting Fermi system described by the Luttinger model~\cite{Giamarchi_2003} and brought out of equilibrium by a sudden quench of the interaction. In the bosonization language, the Hamiltonian of the system after the quench can be written in terms of collective LL degrees of freedom $\phi(x),\theta(x)$ as
\be\label{eqn:HLL}
\m{H}=\frac{u}{2\pi}\int dx
\left[K\left(\partial_x\theta(x)\right)^2 + \frac{1}{K}\left(\partial_x \phi(x)\right)^2
\right]
\ee
where the initial state $\vert\Psi_0\rangle$ corresponds to the ground state of Eq~(\ref{eqn:HLL}) with Luttinger parameter $K_0\neq K$
\footnote{Strictly speaking we also change the sound velocity $u_0\neq u$, such that $u_0K_0=uK$. This condition preserves Galilean invariance}. The dynamics after this global quantum quench has been studied in great detail~\cite{Cazalilla_long09}. Since we are interested in the correlator defined in Eq.~(\ref{eqn:Ddef2}) we have to discuss the nature of the local perturbation.  Here we consider a static potential that couples to the electron density through a forward (fs) and a backward (bs) scattering term~\cite{Giamarchi_2003,GogolinNerseyanTsvelik_2004} which in bosonic variables is
\be\label{eqn:Vloc}
V_{\rm loc}\equiv V_{\rm fs}+V_{\rm bs}=g_{\rm fs}\,\partial_x\,\phi(x)\vert_{x=0}+g_{\rm bs}\,\cos\,2\phi(x=0)\,.
\ee
The evaluation of the transient OC correlator $\m{D}(t',t)$ greatly simplifies by noticing that, as in equilibrium, the forward and backward scattering processes are decoupled, i.e. they involve independent degrees of freedom~\cite{SM_ImpurityLL}.
As a result, we find that the OC correlator factorizes into $\m{D}(t',t)=\m{D}_{\rm fs}(t',t)\,\m{D}_{\rm bs}(t',t)$.

\textit{Forward-Scattering - } Let us start by discussing the forward scattering contribution, which can be computed exactly
using a method due to Schotte and Schotte~\cite{SchotteSchottePR69,GogolinNerseyanTsvelik_2004}. The final result is
\be
\m{D}_{\rm fs}(t',t)=\langle T\,e^{-i\eta\theta(0,t')}\,e^{i\eta\theta(0,t)}\rangle
\ee
with $\eta=g_{\rm fs}\,K/u$, which can be evaluated in terms of local correlators of the quenched LL~\cite{SM_ImpurityLL}.
It is useful to write the OC correlator as a function of the variables $t'-t=\tau$ and $t$
\be
D_{\rm fs}(\tau;t)=\frac{e^{-i\varphi(\tau)}}{\left[1+\left(\Lambda\tau\right)^2\right]^{\delta^{\rm oc}_{\rm neq}/2}}\,
f_t(\tau)
\ee
where $\Lambda$ is an ultra-violet cut-off and the transient function $f_t(\tau)$ is
\be\label{eqn:transient}
f_t(\tau)=\left(\frac
{[1+\Lambda^2(2t+\tau)^2]^2}
{[1+(2\Lambda\,t)^2]\,[1+4\Lambda^2(t+\tau)^2]}
\right)^{\delta^{\rm oc}_{\rm tr}/4}
\ee
\begin{figure}[t]
\begin{center}
\epsfig{figure=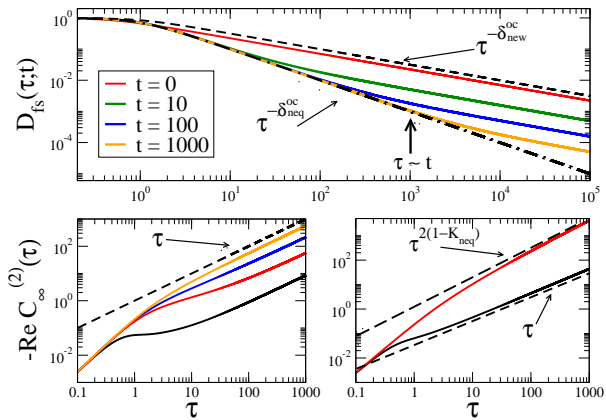,scale=0.32}
\caption{Top Panel: Forward scattering contribution to the OC correlator $D_{\rm fs}(\tau;t)$, for $K_0=4, K=0.1$. The dynamics features two distinct power law regimes, a short time regime with exponent $\delta^{\rm oc}_{\rm neq}$ and a long time one with
exponent $\delta^{\rm oc}_{\rm new}$, with the crossover controlled by the total time $t$. Bottom Panel: Lowest-order back-scattering contribution to the OC correlator, $\mbox{Re} C^{(2)}_{\infty}(\tau)$. In the left panel for $K_0=1.5$ and $K=0.5,0.75.1.0,2.5$ (top to bottom) corresponds to the \emph{thermal} regime, with a linear power law divergence. In the right panel the transition to the strong coupling regime for $K_0=0.75$ and $K=0.25,2.5$ (top to bottom) respectively below and above $K_{\star}\simeq 0.43$.}
\label{fig:fig1}
\end{center}
\end{figure}
the phase $\varphi(\tau)=\frac{g_{\rm fs}^2}{2u^2}\,K\,\mbox{sign}(\tau)\,\arctan(\Lambda\,\tau)$, while the two exponents read, respectively, $\delta^{\rm oc} _{\rm neq/tr}=\frac{g^2_{\rm fs}}{2u^2}\,K_{\rm neq/tr}$ with  $K_{\rm neq/tr}=K_0(1\pm K^2/K_0^2)/2$. We immediately see that for $K=K_0$ the above expression reduces to the well known result for the equilibrium LL,
$D^{\rm eq}_{\rm fs}(\tau)\sim   \left(\frac{1}{\Lambda\tau}\right)^{\delta^{\rm eq}_{\rm oc}}$
with an OC exponent $\delta^{\rm eq}_{\rm oc}=K_0\,g_{\rm fs}^2/2u^2$~\cite{OgawaFurusakiNagaosaPRL92,GogolinPRL93,GogolinNerseyanTsvelik_2004}.
In the case of a bulk quench, $K_0\neq K$ the situation is more interesting.  As we see from figure~\ref{fig:fig1} when both time arguments are longer than the microscopic time scale, i.e. $t,\tau\gg1/\Lambda$,  the OC correlator features two distinct power-law regimes, with a crossover scale set by the total time $t$ after the bulk quench.
The intermediate-time regime, $1/\Lambda\ll\tau\ll t$, when the duration of the local perturbation is much shorter than the one of the global bulk excitation, can be described by a completely dephased non-equilibrium environment which gives a decay $D_{\rm fs}\sim\tau^{-\delta^{\rm oc}_{\rm neq}}$.  While this would be the leading power-law behavior for a strictly infinite waiting time,  the transient nature of the environment results in a different power-law decay at longer times, i.e. for $\tau\gg t$ we have $D_{\rm fs}\sim\alpha_t\,\tau^{-\delta^{\rm oc}_{\rm new}}$, with a prefactor $\alpha_t\sim 1/t^{\delta^{\rm oc}_{\rm tr}/2}$ that depends on the waiting time $t$. Here $\delta^{\rm oc}_{\rm new}= \delta^{\rm oc}_{\rm neq}-\delta^{\rm oc}_{\rm tr}/2=\left(K_{\rm neq}-K_{\rm tr}/2\right)\,g_{\rm fs}^2/2u^2$ can be larger or smaller than the short time OC exponent $\delta^{\rm oc}_{\rm neq}$ depending on the sign of $K_{\rm tr}$, i.e. whether $K \gtrless K_0$.

\textit{Back-Scattering  -} We now consider the backscattering potential and start with a perturbative calculation of $\m{D}_{\rm bs}(t',t)$. Using the linked cluster theorem we may
write 
\be\label{eqn:Cbs}
\m{C}(t',t)\equiv \log\,\m{D}_{\rm bs}(t',t)=\langle\Psi_0\vert\,T\,e^{-i\,\int_t^{t'}\,dt_1\,V_{\rm bs}(t_1)}\vert\Psi_0\rangle_{c}
\ee
where only connected (c) averages contribute, $V_{\rm bs}$ is defined in~(\ref{eqn:Vloc}), and $V_{\rm bs}(t)$
is in the interaction representation of the post-quench global Hamiltonian. Expanding
in powers of $V_{\rm bs}$ we get to lowest order,
\be\label{eqn:C2}
\m{C}^{(2)}(t',t)=-\frac{g_{\rm bs}^2}{2}
\int_t^{t'} dt_1 dt_2
\langle\,T\left[\cos\left(2\phi(t_1)\right)\cos\left(2\phi(t_2)\right)\right]\rangle_{c}
\ee
The result can be evaluated in terms of local correlators of the quenched LL~\cite{SM_ImpurityLL}.
In equilibrium, $K_0=K$, the OC correlator is a function only of $\tau=t'-t$ and the integral~(\ref{eqn:C2}) can be evaluated analytically~\cite{GogolinNerseyanTsvelik_2004}. The result shows that for $K>1$ the correlator goes to a constant at long times, i.e. perturbation theory in $V_{\rm bs}$ is well behaved. In contrast for $K<1$ perturbative correction blows up at long times,  $\Lambda\tau\gg1$, as $\mbox{Re}\,\m{C}^{(2)}(\tau)\sim -\frac{g_{\rm bs}^2}{\Lambda^2}\,(\Lambda\tau)^{2\left(1-K\right)}$, consistent with the result that the backscattering potential is relevant in this regime~\cite{KaneFisherPRB92}.
A crossover time scale $\tau^{\rm eq}_*(K)$  can be extracted by setting $\m{C}^{(2)}(\tau^{\rm eq}_*)\sim 1$, to give $\Lambda\tau^{\rm eq}_*\sim \left(\Lambda/g_{\rm bs}\right)^{1/(1-K)}$. In the strong coupling phase ($K<1$) one can show that at long times ($\tau\gg\tau^{\rm eq}_*$) the correlator $\m{C}^{(2)}(\tau)$  can be resummed into a logarithmic divergence, which gives rise to a universal power law for $\m{D}_{\rm bs}(\tau)\sim 1/\tau^{1/8}$~\cite{GogolinPRL93,KaneMatveevGlazmanPRB94,ProkofevPRB94,FabrizioGogolinPRB95,FurusakiPRB97,
KomnikEggerGogolinPRB97,VonDelftSchoeller_review98}. Finite temperature effects generally change this power law to an exponential decay of the OC correlator~\cite{SM_ImpurityLL}.

\begin{figure}[t]
\begin{center}
\epsfig{figure=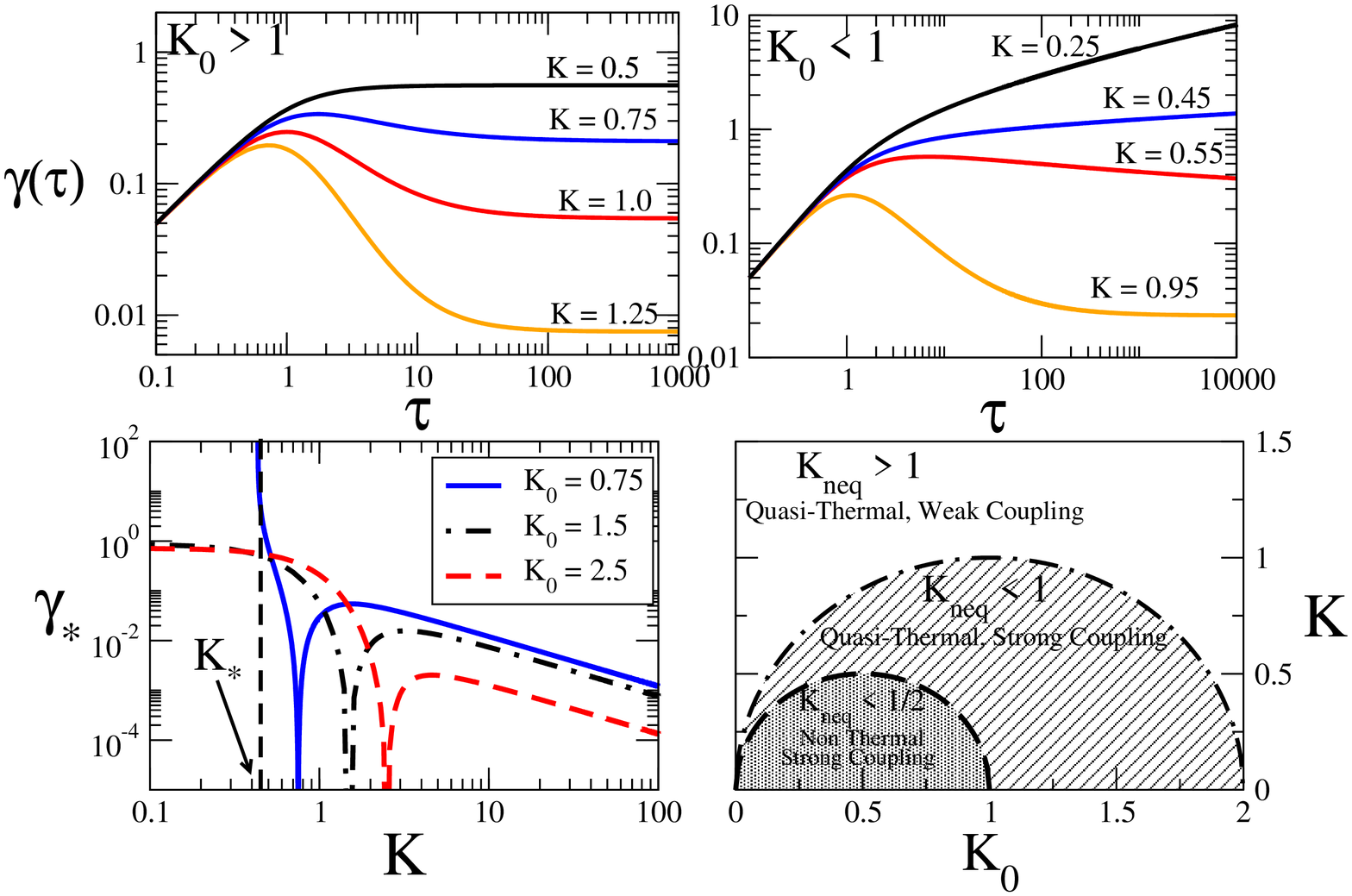,scale=0.32}
\caption{Top Panels: Time dependent relaxation rate $\gamma(\tau)$ for $K_0=1.5$ (left panel) and $K_0=0.75$ (right panel) and different values of $K$. For $K_0<1$ and upon approaching $K\rightarrow K_{\star}=0.43$ the relaxation rate diverges. Bottom Panels: Long time limit of the relaxation rate $\gamma(\tau)\rightarrow\gamma_{\star}$ as a function of $K,K_0$ with $\gamma_{\star}\sim g_{\rm bs}^2(K_0-K)^2$ for $K\rightarrow K_0$
(left panel). Summary of different dynamical regimes (right panel).}
\label{fig:fig2}
\end{center}
\end{figure}

Let us now study the non-equilibrium case, $K_0\neq K$, which is considerably richer. We use the same parametrization as in the forward scattering case, in terms of $\tau=t'-t$ and the total time $t$ after the quench, i.e. $C^{(2)}_t(\tau)$. The integral~(\ref{eqn:C2}) cannot be evaluated in closed analytical form for $K_0\neq K$, yet the structure of the solution can be understood from numerics. To simplify the discussion let us assume first an infinite waiting time after the quench, i.e. $t\rightarrow\infty$, at fixed $\tau$.  In figure~\ref{fig:fig1} we plot $-\mbox{Re}\,\m{C}^{(2)}_{\infty}(\tau)$ which reveals two different behaviors depending on the values of $K_0$ and $K$. For $K_0>1$ or $K_0<1$ and $K>K_{\star}=\sqrt{K_0(1-K_0)}$ (see lower right panel Fig~\ref{fig:fig2}), corresponding to $K_{\rm neq}>1/2$,  the leading long time behavior has a linear divergence $\mbox{Re}\,\m{C}^{(2)}_{\infty}(\tau)\sim -\gamma_{\star}\,\tau$ with a prefactor that can be evaluated in closed form (all energy scales in units of $\Lambda$)
\be
\gamma_{\star}=
g_{\rm bs}^2\left(\frac{2\pi}{2^{K_{\rm neq}}\left[2K_{\rm neq}-1\right]}\right)
\frac{\Gamma\left(2K_{\rm neq}\right)}
{\Gamma\left(K_{\rm neq}+K\right)\Gamma\left(K_{\rm neq}-K\right)}
 \ee
 where $\Gamma(x)$ is the Gamma function.
Such behavior results, through Eq.~(\ref{eqn:Cbs}), into an exponential decay of the OC correlator $\m{D}_{\rm bs}(\tau)\sim \exp\left(-\gamma_*\,\tau\right)$, resulting in a Lorentzian work distribution $P(W)$ of width $\gamma_{\star}$ and reminiscent of an equilibrium finite temperature behavior~\cite{SM_ImpurityLL}.

To further investigate this regime we define a time dependent relaxation rate $\gamma(\tau)=-\partial_{\tau}\,\mbox{Re}\,C^{(2)}_{\infty}(\tau)$ and plot it in figure~\ref{fig:fig2} (top panels) for different values of $K_0,K$.
From this we see that upon approaching $K\rightarrow K_{\star}$ (right panel) the relaxation rate diverges and, quite differently from the thermal equilibrium case, we find a region of parameters for $K_0<1$ and $K<K_{\star}$ (see lower right panel Fig~\ref{fig:fig2}), corresponding to $K_{\rm neq}<1/2$, where the correlator $C^{(2)}_{\infty}(\tau)$ still diverges as a power law at long time but with a new exponent, $\mbox{Re}\,\m{C}^{(2)}_{\infty}(\tau)\sim
-\frac{g_{\rm bs}^2}{\Lambda^2}\,(\Lambda\tau)^{2\left(1-K_{\rm neq}\right)}$. Such a divergence suggests that  the problem retains in this regime a strong coupling nature and allows us to define a non-equilibrium crossover time scale $\Lambda\tau^{\rm neq}_*\sim \left(\Lambda/g_{\rm bs}\right)^{1/(1-K_{\rm neq})}$ controlling the flow of the back-scattering potential (see below).
While these results have been obtained assuming an infinite waiting time after the quench, finite time effects do not seem to qualitatively change this behavior~\cite{SM_ImpurityLL}.

We now consider the effect of higher order backscattering terms
using a time dependent Renormalization Group method recently developed for the bulk Sine-Gordon problem~\cite{Aditi_PRL12}. We refer the reader to Ref.~\onlinecite{SM_ImpurityLL} for details about the derivation of the RG flow and here discuss the main results. At two-loop order we obtain renormalization corrections to the vertex $g_{\rm bs}$ as well as to the quadratic part of the action, that we parameterize in terms of a dissipation $\eta$ (an effective friction for the local coordinate due to its coupling to the bulk
degrees of freedom) and an effective temperature $T_{\rm eff}$~\cite{MitraGiamarchiPRL11,SM_ImpurityLL}. Setting $\Lambda=1$ we obtain flow equations,
\begin{eqnarray}
\frac{dg_{\rm bs}}{d\ln{l}} = g_{\rm bs}\left[1-\left(K_{\rm neq} + \frac{K_{\rm tr}}{1+4 T_m^2}\right)\right]
\label{rg1}\\
\frac{d\eta}{d\ln{l}} = 2\,g_{\rm bs}^2\,I_{\eta}(T_m)\label{rg2}\\
\frac{d(\eta T_{\rm eff})}{d\ln{l}} = \eta T_{\rm eff} + g_{\rm bs}^2\,I_{T_{\rm eff}}(T_m); \frac{dT_m}{d\ln{l}}=-T_m
\end{eqnarray}
where $T_m$ is the time after the quench, and $I_{\eta}(T_m)$, $I_{T_{\rm eff}}(T_m)$ are given in Ref.~\onlinecite{SM_ImpurityLL}. The initial conditions for the RG are $\eta(l=1)=2/(\pi K), T_{\rm eff}(l=1)=0$. If we take the long time ($T_m\rightarrow\infty$) limit, from the RG flow we immediately see that (i) for $K_{\rm neq}>1$ the back-scattering is irrelevant and yet it generates an effective temperature whose value is $T_{\rm eff}\sim g_{\rm bs}^2I_{T_{\rm eff}}(T_m=\infty)/\eta(l=1)$, (ii) for $K_{\rm neq}<1$ the problem flows to strong coupling on time scales larger than $\tau^{\rm neq}_*$, yet a well defined effective temperature can still be identified, at least as long as the relaxation rate $\gamma_*$ or the local dissipation $\eta$, stays finite, i.e. for $K_{\rm neq}>1/2$  (see figure~\ref{fig:fig2}). Deep in the strong coupling phase the perturbative analysis suggests that the OC correlator might eventually keep its power law behavior, a result which would be remarkable, yet from the current analysis we cannot firmly conclude whether a genuine non-equilibrium strong coupling regime remains intact or higher order corrections eventually cut the RG flow.

We note that the effective-temperature $T_{\rm eff}$ is not equivalent to a true temperature $T$, since the latter implies a relaxation rate in the OC correlator  $\gamma_{T}\propto g^2_{\rm bs} T^{2K -1}$~\cite{SM_ImpurityLL} while the former gives a relaxation rate $\gamma_* \propto T_{\rm eff}/({2K_{\rm neq}-1})$.
Thus even in the long time limit, and for a local non-linearity, important differences arise between the OC physics in the presence of a thermalized bath and the one studied in this paper, where the bath is in a non-equilibrium prethermalized state.
It is interesting to note that an effective temperature is also generated in very different contexts such as 
nonequilibrium steady-states arising due to externally imposed voltage bias~\cite{MitraMillisPRB07} and noise~\cite{DallaTorreetalPRB2012}.

\textit{Conclusions -} We have introduced a non-equilibrium OC problem, highlighting its experimental relevance
and have studied it in detail for 1D interacting fermions out of equilibrium due to a global quench.
We have discussed how the non-equilibrium environment
affects the forward and backward scattering contribution to the OC correlator. We have shown that, while the former retains its power law structure even in the presence of a strong non-equilibrium perturbation
the latter is a source of non-linearity which generates locally an exponential decay in time of the OC correlator.
Such a result appears to be consistent with a recent numerical study on Loschmidt echo decay in a 1D spin chain after a local quench starting 
from a highly excited state~\cite{Santos_arxiv14}.

We expect the physics of this \emph{quench-induced} decoherence
to be relevant in other contexts as well, most notably in steady-state transport, where it will result in a non-vanishing zero bias conductance in the weak-link limit as well as a non-zero backscattering correction in the dual limit~\cite{SchiroMitra_Inprep}, a result which is consistent with the non-vanishing impurity density of states recently found in~\cite{KennesMedenPRB13}.

\textit{Acknowledgment - } It is a pleasure to thank D. Huse, A. Rosch, M. Fabrizio, A. Silva, A. Gambassi and L. Santos for helpful discussions. This work was partially supported by a grant from the Simons Foundation (AM)
and by the National Science Foundation under Grants No. PHY 11-25915 (MS and AM) and No. DMR-1004589 (AM).



\begin{thebibliography}{61}
\expandafter\ifx\csname natexlab\endcsname\relax\def\natexlab#1{#1}\fi
\expandafter\ifx\csname bibnamefont\endcsname\relax
  \def\bibnamefont#1{#1}\fi
\expandafter\ifx\csname bibfnamefont\endcsname\relax
  \def\bibfnamefont#1{#1}\fi
\expandafter\ifx\csname citenamefont\endcsname\relax
  \def\citenamefont#1{#1}\fi
\expandafter\ifx\csname url\endcsname\relax
  \def\url#1{\texttt{#1}}\fi
\expandafter\ifx\csname urlprefix\endcsname\relax\def\urlprefix{URL }\fi
\providecommand{\bibinfo}[2]{#2}
\providecommand{\eprint}[2][]{\url{#2}}

\bibitem[{\citenamefont{Anderson}(1967)}]{Anderson_prl67}
\bibinfo{author}{\bibfnamefont{P.~W.} \bibnamefont{Anderson}},
  \bibinfo{journal}{Phys. Rev. Lett.} \textbf{\bibinfo{volume}{18}},
  \bibinfo{pages}{1049} (\bibinfo{year}{1967}).

\bibitem[{\citenamefont{Nozi\'eres and De~Domincis}(1969)}]{XrayEdgeND}
\bibinfo{author}{\bibfnamefont{P.}~\bibnamefont{Nozi\'eres}} \bibnamefont{and}
  \bibinfo{author}{\bibfnamefont{C.~T.} \bibnamefont{De~Domincis}},
  \bibinfo{journal}{Phys. Rev.} \textbf{\bibinfo{volume}{178}},
  \bibinfo{pages}{1097} (\bibinfo{year}{1969}).

\bibitem[{\citenamefont{Gogolin}(1993)}]{GogolinPRL93}
\bibinfo{author}{\bibfnamefont{A.~O.} \bibnamefont{Gogolin}},
  \bibinfo{journal}{Phys. Rev. Lett.} \textbf{\bibinfo{volume}{71}},
  \bibinfo{pages}{2995} (\bibinfo{year}{1993}).

\bibitem[{\citenamefont{Kane and Fisher}(1992{\natexlab{a}})}]{KaneFisherPRL92}
\bibinfo{author}{\bibfnamefont{C.~L.} \bibnamefont{Kane}} \bibnamefont{and}
  \bibinfo{author}{\bibfnamefont{M.~P.~A.} \bibnamefont{Fisher}},
  \bibinfo{journal}{Phys. Rev. Lett.} \textbf{\bibinfo{volume}{68}},
  \bibinfo{pages}{1220} (\bibinfo{year}{1992}{\natexlab{a}}).

\bibitem[{\citenamefont{Kane and Fisher}(1992{\natexlab{b}})}]{KaneFisherPRB92}
\bibinfo{author}{\bibfnamefont{C.~L.} \bibnamefont{Kane}} \bibnamefont{and}
  \bibinfo{author}{\bibfnamefont{M.~P.~A.} \bibnamefont{Fisher}},
  \bibinfo{journal}{Phys. Rev. B} \textbf{\bibinfo{volume}{46}},
  \bibinfo{pages}{15233} (\bibinfo{year}{1992}{\natexlab{b}}).

\bibitem[{\citenamefont{Meden et~al.}(1998)\citenamefont{Meden, Schmitteckert,
  and Shannon}}]{MedenEtAlPRB98}
\bibinfo{author}{\bibfnamefont{V.}~\bibnamefont{Meden}},
  \bibinfo{author}{\bibfnamefont{P.}~\bibnamefont{Schmitteckert}},
  \bibnamefont{and} \bibinfo{author}{\bibfnamefont{N.}~\bibnamefont{Shannon}},
  \bibinfo{journal}{Phys. Rev. B} \textbf{\bibinfo{volume}{57}},
  \bibinfo{pages}{8878} (\bibinfo{year}{1998}).

\bibitem[{\citenamefont{Yuval and Anderson}(1970)}]{AY}
\bibinfo{author}{\bibfnamefont{G.}~\bibnamefont{Yuval}} \bibnamefont{and}
  \bibinfo{author}{\bibfnamefont{P.~W.} \bibnamefont{Anderson}},
  \bibinfo{journal}{Phys. Rev. B} \textbf{\bibinfo{volume}{1}},
  \bibinfo{pages}{1522} (\bibinfo{year}{1970}).

\bibitem[{\citenamefont{Anderson et~al.}(1970)\citenamefont{Anderson, Yuval,
  and Hamann}}]{AY2}
\bibinfo{author}{\bibfnamefont{P.~W.} \bibnamefont{Anderson}},
  \bibinfo{author}{\bibfnamefont{G.}~\bibnamefont{Yuval}}, \bibnamefont{and}
  \bibinfo{author}{\bibfnamefont{D.~R.} \bibnamefont{Hamann}},
  \bibinfo{journal}{Phys. Rev. B} \textbf{\bibinfo{volume}{1}},
  \bibinfo{pages}{4464} (\bibinfo{year}{1970}).

\bibitem[{\citenamefont{T\"ureci et~al.}(2011)\citenamefont{T\"ureci, Hanl,
  Claassen, Weichselbaum, Hecht, Braunecker, Govorov, Glazman, Imamoglu, and
  von Delft}}]{Tureci_prl11}
\bibinfo{author}{\bibfnamefont{H.~E.} \bibnamefont{T\"ureci}},
  \bibinfo{author}{\bibfnamefont{M.}~\bibnamefont{Hanl}},
  \bibinfo{author}{\bibfnamefont{M.}~\bibnamefont{Claassen}},
  \bibinfo{author}{\bibfnamefont{A.}~\bibnamefont{Weichselbaum}},
  \bibinfo{author}{\bibfnamefont{T.}~\bibnamefont{Hecht}},
  \bibinfo{author}{\bibfnamefont{B.}~\bibnamefont{Braunecker}},
  \bibinfo{author}{\bibfnamefont{A.}~\bibnamefont{Govorov}},
  \bibinfo{author}{\bibfnamefont{L.}~\bibnamefont{Glazman}},
  \bibinfo{author}{\bibfnamefont{A.}~\bibnamefont{Imamoglu}}, \bibnamefont{and}
  \bibinfo{author}{\bibfnamefont{J.}~\bibnamefont{von Delft}},
  \bibinfo{journal}{Phys. Rev. Lett.} \textbf{\bibinfo{volume}{106}},
  \bibinfo{pages}{107402} (\bibinfo{year}{2011}).

\bibitem[{\citenamefont{Latta et~al.}(2011)\citenamefont{Latta, Haupt, Hanl,
  Weichselbaum, Claassen, Wuester, Fallahi, Faelt, Glazman, Von~Delft
  et~al.}}]{Imamoglu_nature11}
\bibinfo{author}{\bibfnamefont{C.}~\bibnamefont{Latta}},
  \bibinfo{author}{\bibfnamefont{F.}~\bibnamefont{Haupt}},
  \bibinfo{author}{\bibfnamefont{M.}~\bibnamefont{Hanl}},
  \bibinfo{author}{\bibfnamefont{A.}~\bibnamefont{Weichselbaum}},
  \bibinfo{author}{\bibfnamefont{M.}~\bibnamefont{Claassen}},
  \bibinfo{author}{\bibfnamefont{W.}~\bibnamefont{Wuester}},
  \bibinfo{author}{\bibfnamefont{P.}~\bibnamefont{Fallahi}},
  \bibinfo{author}{\bibfnamefont{S.}~\bibnamefont{Faelt}},
  \bibinfo{author}{\bibfnamefont{L.}~\bibnamefont{Glazman}},
  \bibinfo{author}{\bibfnamefont{J.}~\bibnamefont{Von~Delft}},
  \bibnamefont{et~al.}, \bibinfo{journal}{Nature}
  \textbf{\bibinfo{volume}{474}}, \bibinfo{pages}{627} (\bibinfo{year}{2011}).

\bibitem[{\citenamefont{M\"under et~al.}(2012)\citenamefont{M\"under,
  Weichselbaum, Goldstein, Gefen, and von Delft}}]{GoldsteinPRB12}
\bibinfo{author}{\bibfnamefont{W.}~\bibnamefont{M\"under}},
  \bibinfo{author}{\bibfnamefont{A.}~\bibnamefont{Weichselbaum}},
  \bibinfo{author}{\bibfnamefont{M.}~\bibnamefont{Goldstein}},
  \bibinfo{author}{\bibfnamefont{Y.}~\bibnamefont{Gefen}}, \bibnamefont{and}
  \bibinfo{author}{\bibfnamefont{J.}~\bibnamefont{von Delft}},
  \bibinfo{journal}{Phys. Rev. B} \textbf{\bibinfo{volume}{85}},
  \bibinfo{pages}{235104} (\bibinfo{year}{2012}).

\bibitem[{\citenamefont{Bloch et~al.}(2008)\citenamefont{Bloch, Dalibard, and
  Zwerger}}]{Bloch_rmp08}
\bibinfo{author}{\bibfnamefont{I.}~\bibnamefont{Bloch}},
  \bibinfo{author}{\bibfnamefont{J.}~\bibnamefont{Dalibard}}, \bibnamefont{and}
  \bibinfo{author}{\bibfnamefont{W.}~\bibnamefont{Zwerger}},
  \bibinfo{journal}{Rev. Mod. Phys.} \textbf{\bibinfo{volume}{80}},
  \bibinfo{pages}{885} (\bibinfo{year}{2008}).

\bibitem[{\citenamefont{Weitenberg et~al.}(2011)\citenamefont{Weitenberg,
  Endres, Scherson, Cheneau, Schausz, Fukuhara, Bloch, and
  Kuhr}}]{Weitenberg11}
\bibinfo{author}{\bibfnamefont{C.}~\bibnamefont{Weitenberg}},
  \bibinfo{author}{\bibfnamefont{M.}~\bibnamefont{Endres}},
  \bibinfo{author}{\bibfnamefont{J.}~\bibnamefont{Scherson}},
  \bibinfo{author}{\bibfnamefont{M.}~\bibnamefont{Cheneau}},
  \bibinfo{author}{\bibfnamefont{P.}~\bibnamefont{Schausz}},
  \bibinfo{author}{\bibfnamefont{T.}~\bibnamefont{Fukuhara}},
  \bibinfo{author}{\bibfnamefont{I.}~\bibnamefont{Bloch}}, \bibnamefont{and}
  \bibinfo{author}{\bibfnamefont{S.}~\bibnamefont{Kuhr}},
  \bibinfo{journal}{Nature} \textbf{\bibinfo{volume}{471}},
  \bibinfo{pages}{319} (\bibinfo{year}{2011}).

\bibitem[{\citenamefont{Fukuhara et~al.}(2013)\citenamefont{Fukuhara, Kantian,
  Endres, Cheneau, Schausz, Hild, Bellem, Schollwock, Giamarchi, Gross
  et~al.}}]{FukuharaNatPhys13}
\bibinfo{author}{\bibfnamefont{T.}~\bibnamefont{Fukuhara}},
  \bibinfo{author}{\bibfnamefont{A.}~\bibnamefont{Kantian}},
  \bibinfo{author}{\bibfnamefont{M.}~\bibnamefont{Endres}},
  \bibinfo{author}{\bibfnamefont{M.}~\bibnamefont{Cheneau}},
  \bibinfo{author}{\bibfnamefont{P.}~\bibnamefont{Schausz}},
  \bibinfo{author}{\bibfnamefont{S.}~\bibnamefont{Hild}},
  \bibinfo{author}{\bibfnamefont{D.}~\bibnamefont{Bellem}},
  \bibinfo{author}{\bibfnamefont{U.}~\bibnamefont{Schollwock}},
  \bibinfo{author}{\bibfnamefont{T.}~\bibnamefont{Giamarchi}},
  \bibinfo{author}{\bibfnamefont{C.}~\bibnamefont{Gross}},
  \bibnamefont{et~al.}, \bibinfo{journal}{Nature Physics}
  \textbf{\bibinfo{volume}{9}}, \bibinfo{pages}{235} (\bibinfo{year}{2013}).

\bibitem[{\citenamefont{Knap et~al.}(2012)\citenamefont{Knap, Shashi, Nishida,
  Imambekov, Abanin, and Demler}}]{KnapetalPRX12}
\bibinfo{author}{\bibfnamefont{M.}~\bibnamefont{Knap}},
  \bibinfo{author}{\bibfnamefont{A.}~\bibnamefont{Shashi}},
  \bibinfo{author}{\bibfnamefont{Y.}~\bibnamefont{Nishida}},
  \bibinfo{author}{\bibfnamefont{A.}~\bibnamefont{Imambekov}},
  \bibinfo{author}{\bibfnamefont{D.~A.} \bibnamefont{Abanin}},
  \bibnamefont{and} \bibinfo{author}{\bibfnamefont{E.}~\bibnamefont{Demler}},
  \bibinfo{journal}{Phys. Rev. X} \textbf{\bibinfo{volume}{2}},
  \bibinfo{pages}{041020} (\bibinfo{year}{2012}).

\bibitem[{\citenamefont{Ng}(1996)}]{NgPRB96}
\bibinfo{author}{\bibfnamefont{T.-K.} \bibnamefont{Ng}},
  \bibinfo{journal}{Phys. Rev. B} \textbf{\bibinfo{volume}{54}},
  \bibinfo{pages}{5814} (\bibinfo{year}{1996}).

\bibitem[{\citenamefont{Muzykantskii et~al.}(2003)\citenamefont{Muzykantskii,
  d'Ambrumenil, and Braunecker}}]{MuzykantskiiEtAlPRL03}
\bibinfo{author}{\bibfnamefont{B.}~\bibnamefont{Muzykantskii}},
  \bibinfo{author}{\bibfnamefont{N.}~\bibnamefont{d'Ambrumenil}},
  \bibnamefont{and}
  \bibinfo{author}{\bibfnamefont{B.}~\bibnamefont{Braunecker}},
  \bibinfo{journal}{Phys. Rev. Lett.} \textbf{\bibinfo{volume}{91}},
  \bibinfo{pages}{266602} (\bibinfo{year}{2003}).

\bibitem[{\citenamefont{Braunecker}(2003)}]{BrauneckerPRB03}
\bibinfo{author}{\bibfnamefont{B.}~\bibnamefont{Braunecker}},
  \bibinfo{journal}{Phys. Rev. B} \textbf{\bibinfo{volume}{68}},
  \bibinfo{pages}{153104} (\bibinfo{year}{2003}).

\bibitem[{\citenamefont{Abanin and Levitov}(2005)}]{AbaninLevitovPRL04}
\bibinfo{author}{\bibfnamefont{D.~A.} \bibnamefont{Abanin}} \bibnamefont{and}
  \bibinfo{author}{\bibfnamefont{L.~S.} \bibnamefont{Levitov}},
  \bibinfo{journal}{Phys. Rev. Lett.} \textbf{\bibinfo{volume}{94}},
  \bibinfo{pages}{186803} (\bibinfo{year}{2005}).

\bibitem[{\citenamefont{Mitra and Millis}(2007)}]{MitraMillisPRB07}
\bibinfo{author}{\bibfnamefont{A.}~\bibnamefont{Mitra}} \bibnamefont{and}
  \bibinfo{author}{\bibfnamefont{A.~J.} \bibnamefont{Millis}},
  \bibinfo{journal}{Phys. Rev. B} \textbf{\bibinfo{volume}{76}},
  \bibinfo{pages}{085342} (\bibinfo{year}{2007}).

\bibitem[{\citenamefont{Segal et~al.}(2007)\citenamefont{Segal, Reichman, and
  Millis}}]{SegalPRB07}
\bibinfo{author}{\bibfnamefont{D.}~\bibnamefont{Segal}},
  \bibinfo{author}{\bibfnamefont{D.~R.} \bibnamefont{Reichman}},
  \bibnamefont{and} \bibinfo{author}{\bibfnamefont{A.~J.}
  \bibnamefont{Millis}}, \bibinfo{journal}{Phys. Rev. B}
  \textbf{\bibinfo{volume}{76}}, \bibinfo{pages}{195316}
  (\bibinfo{year}{2007}).

\bibitem[{\citenamefont{Dalla~Torre et~al.}(2012)\citenamefont{Dalla~Torre,
  Demler, Giamarchi, and Altman}}]{DallaTorreetalPRB2012}
\bibinfo{author}{\bibfnamefont{E.~G.} \bibnamefont{Dalla~Torre}},
  \bibinfo{author}{\bibfnamefont{E.}~\bibnamefont{Demler}},
  \bibinfo{author}{\bibfnamefont{T.}~\bibnamefont{Giamarchi}},
  \bibnamefont{and} \bibinfo{author}{\bibfnamefont{E.}~\bibnamefont{Altman}},
  \bibinfo{journal}{Phys. Rev. B} \textbf{\bibinfo{volume}{85}},
  \bibinfo{pages}{184302} (\bibinfo{year}{2012}).

\bibitem[{\citenamefont{Berges et~al.}(2004)\citenamefont{Berges, Bors\'anyi,
  and Wetterich}}]{Berges04}
\bibinfo{author}{\bibfnamefont{J.}~\bibnamefont{Berges}},
  \bibinfo{author}{\bibfnamefont{S.}~\bibnamefont{Bors\'anyi}},
  \bibnamefont{and}
  \bibinfo{author}{\bibfnamefont{C.}~\bibnamefont{Wetterich}},
  \bibinfo{journal}{Phys. Rev. Lett.} \textbf{\bibinfo{volume}{93}},
  \bibinfo{pages}{142002} (\bibinfo{year}{2004}).

\bibitem[{\citenamefont{Moeckel and Kehrein}(2008)}]{Kehrein_prl08}
\bibinfo{author}{\bibfnamefont{M.}~\bibnamefont{Moeckel}} \bibnamefont{and}
  \bibinfo{author}{\bibfnamefont{S.}~\bibnamefont{Kehrein}},
  \bibinfo{journal}{Phys. Rev. Lett.} \textbf{\bibinfo{volume}{100}},
  \bibinfo{pages}{175702} (\bibinfo{year}{2008}).

\bibitem[{\citenamefont{Gring et~al.}(2012)\citenamefont{Gring, Kuhnert,
  Langen, Kitagawa, Rauer, Schreitl, Mazets, Smith, Demler, and
  Schmiedmayer}}]{GringScience12}
\bibinfo{author}{\bibfnamefont{M.}~\bibnamefont{Gring}},
  \bibinfo{author}{\bibfnamefont{M.}~\bibnamefont{Kuhnert}},
  \bibinfo{author}{\bibfnamefont{T.}~\bibnamefont{Langen}},
  \bibinfo{author}{\bibfnamefont{T.}~\bibnamefont{Kitagawa}},
  \bibinfo{author}{\bibfnamefont{B.}~\bibnamefont{Rauer}},
  \bibinfo{author}{\bibfnamefont{M.}~\bibnamefont{Schreitl}},
  \bibinfo{author}{\bibfnamefont{I.}~\bibnamefont{Mazets}},
  \bibinfo{author}{\bibfnamefont{D.~A.} \bibnamefont{Smith}},
  \bibinfo{author}{\bibfnamefont{E.}~\bibnamefont{Demler}}, \bibnamefont{and}
  \bibinfo{author}{\bibfnamefont{J.}~\bibnamefont{Schmiedmayer}},
  \bibinfo{journal}{Science} \textbf{\bibinfo{volume}{337}},
  \bibinfo{pages}{1318} (\bibinfo{year}{2012}).

\bibitem[{\citenamefont{Karrasch et~al.}(2012)\citenamefont{Karrasch, Rentrop,
  Schuricht, and Meden}}]{Karrasch_PRL12}
\bibinfo{author}{\bibfnamefont{C.}~\bibnamefont{Karrasch}},
  \bibinfo{author}{\bibfnamefont{J.}~\bibnamefont{Rentrop}},
  \bibinfo{author}{\bibfnamefont{D.}~\bibnamefont{Schuricht}},
  \bibnamefont{and} \bibinfo{author}{\bibfnamefont{V.}~\bibnamefont{Meden}},
  \bibinfo{journal}{Phys. Rev. Lett.} \textbf{\bibinfo{volume}{109}},
  \bibinfo{pages}{126406} (\bibinfo{year}{2012}).

\bibitem[{\citenamefont{Mitra}(2013)}]{MitraPRB13}
\bibinfo{author}{\bibfnamefont{A.}~\bibnamefont{Mitra}},
  \bibinfo{journal}{Phys. Rev. B} \textbf{\bibinfo{volume}{87}},
  \bibinfo{pages}{205109} (\bibinfo{year}{2013}).

\bibitem[{\citenamefont{Peres}(1984)}]{Peres_PRA84}
\bibinfo{author}{\bibfnamefont{A.}~\bibnamefont{Peres}},
  \bibinfo{journal}{Phys. Rev. A} \textbf{\bibinfo{volume}{30}},
  \bibinfo{pages}{1610} (\bibinfo{year}{1984}).

\bibitem[{\citenamefont{Gorin et~al.}(2006)\citenamefont{Gorin, Prosen,
  Seligman, and Žnidarič}}]{Gorin_PhysRep06}
\bibinfo{author}{\bibfnamefont{T.}~\bibnamefont{Gorin}},
  \bibinfo{author}{\bibfnamefont{T.}~\bibnamefont{Prosen}},
  \bibinfo{author}{\bibfnamefont{T.~H.} \bibnamefont{Seligman}},
  \bibnamefont{and}
  \bibinfo{author}{\bibfnamefont{M.}~\bibnamefont{Žnidarič}},
  \bibinfo{journal}{Physics Reports} \textbf{\bibinfo{volume}{435}},
  \bibinfo{pages}{33 } (\bibinfo{year}{2006}), ISSN \bibinfo{issn}{0370-1573}.

\bibitem[{\citenamefont{Heyl and
  Kehrein}(2012{\natexlab{a}})}]{HeylKehreinPRB12}
\bibinfo{author}{\bibfnamefont{M.}~\bibnamefont{Heyl}} \bibnamefont{and}
  \bibinfo{author}{\bibfnamefont{S.}~\bibnamefont{Kehrein}},
  \bibinfo{journal}{Phys. Rev. B} \textbf{\bibinfo{volume}{85}},
  \bibinfo{pages}{155413} (\bibinfo{year}{2012}{\natexlab{a}}).

\bibitem[{\citenamefont{D\'ora et~al.}(2013)\citenamefont{D\'ora, Pollmann,
  Fort\'agh, and Zar\'and}}]{DoraetalPRL2013}
\bibinfo{author}{\bibfnamefont{B.}~\bibnamefont{D\'ora}},
  \bibinfo{author}{\bibfnamefont{F.}~\bibnamefont{Pollmann}},
  \bibinfo{author}{\bibfnamefont{J.}~\bibnamefont{Fort\'agh}},
  \bibnamefont{and} \bibinfo{author}{\bibfnamefont{G.}~\bibnamefont{Zar\'and}},
  \bibinfo{journal}{Phys. Rev. Lett.} \textbf{\bibinfo{volume}{111}},
  \bibinfo{pages}{046402} (\bibinfo{year}{2013}).

\bibitem[{\citenamefont{Silva}(2008)}]{Silva_work_statistics}
\bibinfo{author}{\bibfnamefont{A.}~\bibnamefont{Silva}},
  \bibinfo{journal}{Phys. Rev. Lett.} \textbf{\bibinfo{volume}{101}},
  \bibinfo{pages}{120603} (\bibinfo{year}{2008}).

\bibitem[{\citenamefont{Gambassi and Silva}(2012)}]{GambassiSilvaPRL12}
\bibinfo{author}{\bibfnamefont{A.}~\bibnamefont{Gambassi}} \bibnamefont{and}
  \bibinfo{author}{\bibfnamefont{A.}~\bibnamefont{Silva}},
  \bibinfo{journal}{Phys. Rev. Lett.} \textbf{\bibinfo{volume}{109}},
  \bibinfo{pages}{250602} (\bibinfo{year}{2012}).

\bibitem[{\citenamefont{Heyl and
  Kehrein}(2012{\natexlab{b}})}]{HeylKehreinPRL12}
\bibinfo{author}{\bibfnamefont{M.}~\bibnamefont{Heyl}} \bibnamefont{and}
  \bibinfo{author}{\bibfnamefont{S.}~\bibnamefont{Kehrein}},
  \bibinfo{journal}{Phys. Rev. Lett.} \textbf{\bibinfo{volume}{108}},
  \bibinfo{pages}{190601} (\bibinfo{year}{2012}{\natexlab{b}}).

\bibitem[{\citenamefont{Heyl et~al.}(2013)\citenamefont{Heyl, Polkovnikov, and
  Kehrein}}]{HeylKehreinPolkovnikovPRL13}
\bibinfo{author}{\bibfnamefont{M.}~\bibnamefont{Heyl}},
  \bibinfo{author}{\bibfnamefont{A.}~\bibnamefont{Polkovnikov}},
  \bibnamefont{and} \bibinfo{author}{\bibfnamefont{S.}~\bibnamefont{Kehrein}},
  \bibinfo{journal}{Phys. Rev. Lett.} \textbf{\bibinfo{volume}{110}},
  \bibinfo{pages}{135704} (\bibinfo{year}{2013}).

\bibitem[{\citenamefont{Vasseur et~al.}(2013)\citenamefont{Vasseur, Trinh,
  Haas, and Saleur}}]{Vasseur_etalPRL13}
\bibinfo{author}{\bibfnamefont{R.}~\bibnamefont{Vasseur}},
  \bibinfo{author}{\bibfnamefont{K.}~\bibnamefont{Trinh}},
  \bibinfo{author}{\bibfnamefont{S.}~\bibnamefont{Haas}}, \bibnamefont{and}
  \bibinfo{author}{\bibfnamefont{H.}~\bibnamefont{Saleur}},
  \bibinfo{journal}{Phys. Rev. Lett.} \textbf{\bibinfo{volume}{110}},
  \bibinfo{pages}{240601} (\bibinfo{year}{2013}).

\bibitem[{SM_()}]{SM_ImpurityLL}
\bibinfo{journal}{See supplementary material for further technical details.}

\bibitem[{\citenamefont{{Hickey} and {Genway}}(2014)}]{HickeyGenway_arxiv14}
\bibinfo{author}{\bibfnamefont{J.~M.} \bibnamefont{{Hickey}}} \bibnamefont{and}
  \bibinfo{author}{\bibfnamefont{S.}~\bibnamefont{{Genway}}},
  \bibinfo{journal}{ArXiv e-prints}  (\bibinfo{year}{2014}),
  \eprint{1403.3542}.

\bibitem[{\citenamefont{{Palmai} and
  {Sotiriadis}}(2014)}]{PalmaiSotiriadis_arxiv14}
\bibinfo{author}{\bibfnamefont{T.}~\bibnamefont{{Palmai}}} \bibnamefont{and}
  \bibinfo{author}{\bibfnamefont{S.}~\bibnamefont{{Sotiriadis}}},
  \bibinfo{journal}{ArXiv e-prints}  (\bibinfo{year}{2014}),
  \eprint{1403.7450}.

\bibitem[{\citenamefont{Micheli et~al.}(2004)\citenamefont{Micheli, Daley,
  Jaksch, and Zoller}}]{MicheliEtAlPRL04}
\bibinfo{author}{\bibfnamefont{A.}~\bibnamefont{Micheli}},
  \bibinfo{author}{\bibfnamefont{A.~J.} \bibnamefont{Daley}},
  \bibinfo{author}{\bibfnamefont{D.}~\bibnamefont{Jaksch}}, \bibnamefont{and}
  \bibinfo{author}{\bibfnamefont{P.}~\bibnamefont{Zoller}},
  \bibinfo{journal}{Phys. Rev. Lett.} \textbf{\bibinfo{volume}{93}},
  \bibinfo{pages}{140408} (\bibinfo{year}{2004}).

\bibitem[{\citenamefont{Goold et~al.}(2011)\citenamefont{Goold, Fogarty,
  Lo~Gullo, Paternostro, and Busch}}]{GooldEtAlPRA11}
\bibinfo{author}{\bibfnamefont{J.}~\bibnamefont{Goold}},
  \bibinfo{author}{\bibfnamefont{T.}~\bibnamefont{Fogarty}},
  \bibinfo{author}{\bibfnamefont{N.}~\bibnamefont{Lo~Gullo}},
  \bibinfo{author}{\bibfnamefont{M.}~\bibnamefont{Paternostro}},
  \bibnamefont{and} \bibinfo{author}{\bibfnamefont{T.}~\bibnamefont{Busch}},
  \bibinfo{journal}{Phys. Rev. A} \textbf{\bibinfo{volume}{84}},
  \bibinfo{pages}{063632} (\bibinfo{year}{2011}).

\bibitem[{\citenamefont{Sindona et~al.}(2013)\citenamefont{Sindona, Goold,
  Lo~Gullo, Lorenzo, and Plastina}}]{SindonaEtAlPRL13}
\bibinfo{author}{\bibfnamefont{A.}~\bibnamefont{Sindona}},
  \bibinfo{author}{\bibfnamefont{J.}~\bibnamefont{Goold}},
  \bibinfo{author}{\bibfnamefont{N.}~\bibnamefont{Lo~Gullo}},
  \bibinfo{author}{\bibfnamefont{S.}~\bibnamefont{Lorenzo}}, \bibnamefont{and}
  \bibinfo{author}{\bibfnamefont{F.}~\bibnamefont{Plastina}},
  \bibinfo{journal}{Phys. Rev. Lett.} \textbf{\bibinfo{volume}{111}},
  \bibinfo{pages}{165303} (\bibinfo{year}{2013}).

\bibitem[{\citenamefont{Knap et~al.}(2013)\citenamefont{Knap, Kantian,
  Giamarchi, Bloch, Lukin, and Demler}}]{KnapEtAlPRL13}
\bibinfo{author}{\bibfnamefont{M.}~\bibnamefont{Knap}},
  \bibinfo{author}{\bibfnamefont{A.}~\bibnamefont{Kantian}},
  \bibinfo{author}{\bibfnamefont{T.}~\bibnamefont{Giamarchi}},
  \bibinfo{author}{\bibfnamefont{I.}~\bibnamefont{Bloch}},
  \bibinfo{author}{\bibfnamefont{M.~D.} \bibnamefont{Lukin}}, \bibnamefont{and}
  \bibinfo{author}{\bibfnamefont{E.}~\bibnamefont{Demler}},
  \bibinfo{journal}{Phys. Rev. Lett.} \textbf{\bibinfo{volume}{111}},
  \bibinfo{pages}{147205} (\bibinfo{year}{2013}).

\bibitem[{\citenamefont{Mazzola et~al.}(2013)\citenamefont{Mazzola, De~Chiara,
  and Paternostro}}]{PaternostroPRL13}
\bibinfo{author}{\bibfnamefont{L.}~\bibnamefont{Mazzola}},
  \bibinfo{author}{\bibfnamefont{G.}~\bibnamefont{De~Chiara}},
  \bibnamefont{and}
  \bibinfo{author}{\bibfnamefont{M.}~\bibnamefont{Paternostro}},
  \bibinfo{journal}{Phys. Rev. Lett.} \textbf{\bibinfo{volume}{110}},
  \bibinfo{pages}{230602} (\bibinfo{year}{2013}).

\bibitem[{\citenamefont{Dorner et~al.}(2013)\citenamefont{Dorner, Clark,
  Heaney, Fazio, Goold, and Vedral}}]{FazioPRL13}
\bibinfo{author}{\bibfnamefont{R.}~\bibnamefont{Dorner}},
  \bibinfo{author}{\bibfnamefont{S.~R.} \bibnamefont{Clark}},
  \bibinfo{author}{\bibfnamefont{L.}~\bibnamefont{Heaney}},
  \bibinfo{author}{\bibfnamefont{R.}~\bibnamefont{Fazio}},
  \bibinfo{author}{\bibfnamefont{J.}~\bibnamefont{Goold}}, \bibnamefont{and}
  \bibinfo{author}{\bibfnamefont{V.}~\bibnamefont{Vedral}},
  \bibinfo{journal}{Phys. Rev. Lett.} \textbf{\bibinfo{volume}{110}},
  \bibinfo{pages}{230601} (\bibinfo{year}{2013}).

\bibitem[{\citenamefont{Giamarchi}(2003)}]{Giamarchi_2003}
\bibinfo{author}{\bibfnamefont{T.}~\bibnamefont{Giamarchi}},
  \emph{\bibinfo{title}{Quantum Physics in One Dimension}}
  (\bibinfo{publisher}{Oxford University Press, {USA}}, \bibinfo{year}{2003}),
  \bibinfo{edition}{1st} ed., ISBN \bibinfo{isbn}{0198509146}.

\bibitem[{\citenamefont{Iucci and Cazalilla}(2009)}]{Cazalilla_long09}
\bibinfo{author}{\bibfnamefont{A.}~\bibnamefont{Iucci}} \bibnamefont{and}
  \bibinfo{author}{\bibfnamefont{M.~A.} \bibnamefont{Cazalilla}},
  \bibinfo{journal}{Phys. Rev. A} \textbf{\bibinfo{volume}{80}},
  \bibinfo{pages}{063619} (\bibinfo{year}{2009}).

\bibitem[{\citenamefont{Gogolin et~al.}(2004)\citenamefont{Gogolin, Nersesyan,
  and Tsvelik}}]{GogolinNerseyanTsvelik_2004}
\bibinfo{author}{\bibfnamefont{A.}~\bibnamefont{Gogolin}},
  \bibinfo{author}{\bibfnamefont{A.}~\bibnamefont{Nersesyan}},
  \bibnamefont{and} \bibinfo{author}{\bibfnamefont{A.}~\bibnamefont{Tsvelik}},
  \emph{\bibinfo{title}{Bosonization and Strongly Correlated Systems}}
  (\bibinfo{publisher}{Cambridge University Press, {USA}},
  \bibinfo{year}{2004}), \bibinfo{edition}{1st} ed., ISBN
  \bibinfo{isbn}{0198509146}.

\bibitem[{\citenamefont{Schotte and Schotte}(1969)}]{SchotteSchottePR69}
\bibinfo{author}{\bibfnamefont{K.~D.} \bibnamefont{Schotte}} \bibnamefont{and}
  \bibinfo{author}{\bibfnamefont{U.}~\bibnamefont{Schotte}},
  \bibinfo{journal}{Phys. Rev.} \textbf{\bibinfo{volume}{182}},
  \bibinfo{pages}{479} (\bibinfo{year}{1969}).

\bibitem[{\citenamefont{Ogawa et~al.}(1992)\citenamefont{Ogawa, Furusaki, and
  Nagaosa}}]{OgawaFurusakiNagaosaPRL92}
\bibinfo{author}{\bibfnamefont{T.}~\bibnamefont{Ogawa}},
  \bibinfo{author}{\bibfnamefont{A.}~\bibnamefont{Furusaki}}, \bibnamefont{and}
  \bibinfo{author}{\bibfnamefont{N.}~\bibnamefont{Nagaosa}},
  \bibinfo{journal}{Phys. Rev. Lett.} \textbf{\bibinfo{volume}{68}},
  \bibinfo{pages}{3638} (\bibinfo{year}{1992}).

\bibitem[{\citenamefont{Kane et~al.}(1994)\citenamefont{Kane, Matveev, and
  Glazman}}]{KaneMatveevGlazmanPRB94}
\bibinfo{author}{\bibfnamefont{C.~L.} \bibnamefont{Kane}},
  \bibinfo{author}{\bibfnamefont{K.~A.} \bibnamefont{Matveev}},
  \bibnamefont{and} \bibinfo{author}{\bibfnamefont{L.~I.}
  \bibnamefont{Glazman}}, \bibinfo{journal}{Phys. Rev. B}
  \textbf{\bibinfo{volume}{49}}, \bibinfo{pages}{2253} (\bibinfo{year}{1994}).

\bibitem[{\citenamefont{Prokof'ev}(1994)}]{ProkofevPRB94}
\bibinfo{author}{\bibfnamefont{N.~V.} \bibnamefont{Prokof'ev}},
  \bibinfo{journal}{Phys. Rev. B} \textbf{\bibinfo{volume}{49}},
  \bibinfo{pages}{2148} (\bibinfo{year}{1994}).

\bibitem[{\citenamefont{Fabrizio and Gogolin}(1995)}]{FabrizioGogolinPRB95}
\bibinfo{author}{\bibfnamefont{M.}~\bibnamefont{Fabrizio}} \bibnamefont{and}
  \bibinfo{author}{\bibfnamefont{A.~O.} \bibnamefont{Gogolin}},
  \bibinfo{journal}{Phys. Rev. B} \textbf{\bibinfo{volume}{51}},
  \bibinfo{pages}{17827} (\bibinfo{year}{1995}).

\bibitem[{\citenamefont{Furusaki}(1997)}]{FurusakiPRB97}
\bibinfo{author}{\bibfnamefont{A.}~\bibnamefont{Furusaki}},
  \bibinfo{journal}{Phys. Rev. B} \textbf{\bibinfo{volume}{56}},
  \bibinfo{pages}{9352} (\bibinfo{year}{1997}).

\bibitem[{\citenamefont{Komnik et~al.}(1997)\citenamefont{Komnik, Egger, and
  Gogolin}}]{KomnikEggerGogolinPRB97}
\bibinfo{author}{\bibfnamefont{A.}~\bibnamefont{Komnik}},
  \bibinfo{author}{\bibfnamefont{R.}~\bibnamefont{Egger}}, \bibnamefont{and}
  \bibinfo{author}{\bibfnamefont{A.~O.} \bibnamefont{Gogolin}},
  \bibinfo{journal}{Phys. Rev. B} \textbf{\bibinfo{volume}{56}},
  \bibinfo{pages}{1153} (\bibinfo{year}{1997}).

\bibitem[{\citenamefont{von Delft and
  Schoeller}(1998)}]{VonDelftSchoeller_review98}
\bibinfo{author}{\bibfnamefont{J.}~\bibnamefont{von Delft}} \bibnamefont{and}
  \bibinfo{author}{\bibfnamefont{H.}~\bibnamefont{Schoeller}},
  \bibinfo{journal}{Annalen der Physik} \textbf{\bibinfo{volume}{7}},
  \bibinfo{pages}{225} (\bibinfo{year}{1998}), ISSN \bibinfo{issn}{1521-3889}.

\bibitem[{\citenamefont{Mitra}(2012)}]{Aditi_PRL12}
\bibinfo{author}{\bibfnamefont{A.}~\bibnamefont{Mitra}},
  \bibinfo{journal}{Phys. Rev. Lett.} \textbf{\bibinfo{volume}{109}},
  \bibinfo{pages}{260601} (\bibinfo{year}{2012}).

\bibitem[{\citenamefont{Mitra and Giamarchi}(2011)}]{MitraGiamarchiPRL11}
\bibinfo{author}{\bibfnamefont{A.}~\bibnamefont{Mitra}} \bibnamefont{and}
  \bibinfo{author}{\bibfnamefont{T.}~\bibnamefont{Giamarchi}},
  \bibinfo{journal}{Phys. Rev. Lett.} \textbf{\bibinfo{volume}{107}},
  \bibinfo{pages}{150602} (\bibinfo{year}{2011}).

\bibitem[{\citenamefont{{Torres-Herrera} and {Santos}}(2014)}]{Santos_arxiv14}
\bibinfo{author}{\bibfnamefont{E.~J.} \bibnamefont{{Torres-Herrera}}}
  \bibnamefont{and} \bibinfo{author}{\bibfnamefont{L.~F.}
  \bibnamefont{{Santos}}}, \bibinfo{journal}{ArXiv e-prints}
  (\bibinfo{year}{2014}), \eprint{1402.7084}.

\bibitem[{\citenamefont{Schir\`o and Mitra}(2014)}]{SchiroMitra_Inprep}
\bibinfo{author}{\bibfnamefont{M.}~\bibnamefont{Schir\`o}} \bibnamefont{and}
  \bibinfo{author}{\bibfnamefont{A.}~\bibnamefont{Mitra}}, \bibinfo{journal}{in
  preparation}  (\bibinfo{year}{2014}).

\bibitem[{\citenamefont{Kennes and Meden}(2013)}]{KennesMedenPRB13}
\bibinfo{author}{\bibfnamefont{D.~M.} \bibnamefont{Kennes}} \bibnamefont{and}
  \bibinfo{author}{\bibfnamefont{V.}~\bibnamefont{Meden}},
  \bibinfo{journal}{Phys. Rev. B} \textbf{\bibinfo{volume}{88}},
  \bibinfo{pages}{165131} (\bibinfo{year}{2013}).

\end{thebibliography}

\end{document}


\title{Supplementary Material for "Transient Orthogonality Catastrophe in a Time Dependent Non-Equilibrium Environment"}
\author{M. Schir\'o}
\affiliation{Princeton
 Center for Theoretical Science and Department of Physics, Joseph Henry
  Laboratories, Princeton University, Princeton, NJ
  08544, USA}
  \author{Aditi Mitra}
\affiliation{Department of Physics, New York University, 4 Washington
  Place, New York, New York 10003, USA}
\date{\today}

\maketitle

The supplementary material contains the following: Section I discusses the properties of the transient orthogonality catastrophe (OC) correlator and explicitly shows how it may be measured in experiments. Section II discusses the properties of the quenched Luttinger liquid giving expressions for the Green's functions that are needed for later calculations. Section III shows how the OC correlator 
factorizes into forward and backward scattering components. The OC correlator under the influence of forward scattering is derived.
Section IV discusses the effect of the backward scattering potential on the OC correlator employing perturbation theory. 
Section V treats the effect of the back-scattering potential using a 
time-dependent renormalization group method.

\section{Transient Orthogonality Catastrophe Correlator: General Properties}

In this section we derive a number of results used in the main text for the transient orthogonality catastrophe
(OC) correlator in terms of exact eigenstates of the system. We start from the definition
\be\label{eqn:overlap}
D(\tau =t'-t;t)= \langle\Psi(t')\vert\Psi_{t+}(t')\rangle=\langle\Psi(t)\vert\,e^{i\,\m{H}(t'-t)}\,e^{-i\,\m{H}_+(t'-t)}\vert\Psi(t)\rangle
\ee
where, we recall, $\vert\Psi(t)\rangle=e^{-i\m{H}t}\vert\Psi_0\rangle$, $\vert\Psi_{t+}(t')\rangle\equiv e^{-i\,\m{H}_+(t'-t)}\vert\Psi(t)\rangle$, while $\m{H}_+=\m{H}+V_{loc}$. We then introduce two complete sets of eigenstates
of $\m{H}$ to get
\be
D(\tau ;t)= \sum_{nm}\rho_{nm}(t)\,\langle\Phi_m\vert
e^{i\,\m{H}\tau}\,e^{-i\,\m{H}_+\tau}\vert\Phi_n\rangle
\ee
where $\rho_{nm}(t)=\langle\Phi_n\vert\Psi_0\rangle\langle\Psi_0\vert\Phi_m\rangle\,e^{i\left(E_m-E_n\right)t}$. It is useful to compare this result to the equilibrium finite temperature OC correlator which reads
\be
 D_{eq}(\tau )= \sum_{n}\rho_{n}\,\langle\Phi_n\vert
e^{i\,\m{H}\tau}\,e^{-i\,\m{H}_+\tau}\vert\Phi_n\rangle
\ee
We notice that in equilibrium only the diagonal states contribute with a weight given by their thermal density matrix $\rho_n=e^{-\beta\,E_n}/Z$. We then introduce a complete set of eigenstates
of $\m{H}_+$, $\m{H}_+\vert\tilde{\Phi}_{\alpha}\rangle=\tilde{E}_{\alpha}\vert\tilde{\Phi}_{\alpha}\rangle$ and get
\be
D(\tau; t) = \sum_{nm} \rho_{nm}(t)\,\sum_{\alpha}\,e^{-i\left(\tilde{E}_{\alpha}-E_m\right)\tau}\,
\langle\Phi_m\vert\tilde{\Phi}_{\alpha}\rangle\langle \tilde{\Phi}_{\alpha}\vert\Phi_{n}\rangle
\ee
If we now take the average over the holding time $t$, i.e.
\be
\overline{\rho_{nm}(t)}=\mbox{lim}_{T\rightarrow\infty}\frac{1}{T}\int_0^T\,d\,t\,\rho_{nm}(t)=\delta_{nm}
\langle\,\Phi_n\vert\Psi_0\rangle\langle\Psi_0\vert\Phi_n\rangle=\delta_{nm}\,\rho_{nn}
\ee
we get
\be
\overline{D(\tau ;t)}= \sum_{n} \rho_{nn}\,\sum_{\alpha}\,e^{-i\left(\tilde{E}_{\alpha}-E_n\right)\tau}\,
\vert\langle\Phi_n\vert\tilde{\Phi}_{\alpha}\rangle\vert^2
\ee
This result can be now Fourier transformed to give
\bea
P(W)=\sum_{n\alpha}\,\delta(W-\tilde{E}_\alpha+E_n)\,
\vert\langle\Phi_n\vert\tilde{\Phi}_{\alpha}\rangle\vert^2\,\rho_{nn}
\eea

\subsection{Measuring Transient OC Correlator with Non-Equilibrium Ramsey Interferometry Schemes}

We now turn to discuss a possible experimental protocol to measure the transient OC correlator, based on a non-equilibrium extension of Ramsey interferometry schemes~\cite{MicheliEtAlPRL04,KnapEtAlPRX12,KnapEtAlPRL13}. The basic idea takes advantage of the enormous
progress in building hybrid quantum systems made, for example, by a qubit coupled to a cold atomic gas, and in probing
and manipulating the resulting quantum many body state with single site local resolution. In this spirit, the idea is to use an auxiliary two-level system (TLS) or qubit, coupled to the system through the local perturbation, i.e. to define an auxiliary Hamiltonian
\be\label{eqn:auxiliaryH}
\m{H}[\sigma^z]=\m{H}+V_{\rm loc}\left(1+\sigma^z\right)/2
\ee
living in an extended Hilbert space containing both system and local TLS degrees of freedom. The local TLS can be manipulated using Ramsey pulses which are represented by the following operator acting on the TLS~\cite{KnapEtAlPRL13}
\be
\m{R}\left(\theta,\phi\right)=\frac{1}{\sqrt{2}}
\left[\mathbb{I}\cos(\theta/2)+i\sin(\theta/2)\left(\sigma^+\,e^{i\phi}+
\sigma^-\,e^{-i\phi}\right)\right]
\ee
where $\theta=\Omega_{Rabi}\,T_{pulse}$ while $\phi$ is the phase of the laser. The Ramsey operation acting on the two eigenstates of the TLS, $\vert\uparrow\rangle,\vert\downarrow\rangle$, creates linear superposition states
\bea
\m{R}\left(\theta,\phi\right)\vert\uparrow\rangle=\cos(\theta/2)\vert\uparrow\rangle+i\,e^{-i\phi}\,\sin(\theta/2)\vert\downarrow\rangle
\\
\m{R}\left(\theta,\phi\right)\vert\downarrow\rangle=
\cos(\theta/2)\vert\downarrow\rangle+i\,e^{i\phi}\,\sin(\theta/2)\vert\uparrow\rangle
\eea
Using these operations it is then very natural to design a scheme to manipulate the TLS in such a way as to obtain $D(t';t)$ out of simple local measurements. As an example, we consider the system to be prepared initially at time zero in the product state
\be
\vert\,\Psi_{0\downarrow}\rangle=\vert\Psi_0\rangle\otimes\vert \downarrow\rangle
\ee
We first evolve the system up to some time $t$ with the Hamiltonian $\m{H}[\sigma^z]$ to create the initial non-equilibrium state
\be
e^{-i\m{H}[\sigma^z]\,t}\vert\,\Psi_{0\downarrow}\rangle=\vert\Psi(t)\rangle\otimes\vert \downarrow\rangle
\ee
where as before $\vert\Psi(t)\rangle =e^{-i\m{H}t}\vert\Psi_0\rangle$. We then act with a specific Ramsey pulse on the resulting state, creating a linear superposition for the TLS
\be
\m{R}(\pi/2,-\pi/2)\,e^{-i\m{H}[\sigma^z]\,t}\vert\,\Psi_{0\downarrow}\rangle=
\vert\Psi(t)\rangle\otimes\,\frac{1}{\sqrt{2}}\left(\vert \uparrow\rangle+\vert\downarrow\rangle\right)
\ee
We then evolve again the system with the full Hamiltonian up to time $t'=t+\tau$, with the difference being that now the superposition state triggers nontrivial dynamics which effectively implements the local quench. Indeed one has
\be
e^{-i\m{H}[\sigma^z]\,(t'-t)}\m{R}(\pi/2,-\pi/2)\,e^{-i\m{H}[\sigma^z]\,t}\vert\,\Psi_{0\downarrow}\rangle=
\frac{1}{\sqrt{2}}\left(
e^{-i\m{H}_+(t'-t)}\vert\Psi(t)\rangle\otimes\vert\uparrow\rangle+e^{-i\m{H}(t'-t)}\vert\Psi(t)\rangle\otimes\vert\downarrow\rangle
\right)\equiv \vert\Phi_{t}(t')\rangle
\ee
At the time $t'$ one finally performs a (destructive) measurement of the TLS magnetization along $x,y$ on the state $\vert\Phi_{t}(t')\rangle$ and projects back onto the same state, to give
\bea
\langle\Phi_{t}(t')\vert\,\sigma^x\vert\Phi_{t}(t')\rangle= \mbox{Re}D(t';t)\\
\langle\Phi_{t}(t')\vert\,\sigma^y\vert\Phi_{t}(t')\rangle=-\mbox{Im}D(t';t)
\eea

\subsection{Transient OC Overlap as a Dynamical Correlator of a Local Quantum Degree of Freedom}

We now show that the overlap in Eq.(\ref{eqn:overlap}) can be written in terms of a dynamical correlator of a local quantum degree of freedom, for example a TLS as in previous section or a spinless fermionic level as would be natural in the context of a non-equilibrium X-ray edge problem, coupled to our system through the auxiliary Hamiltonian~(\ref{eqn:auxiliaryH}). The crucial observation is that since $\sigma^z$ is conserved
\be
[\m{H},\sigma^z]=0 
\ee
 the TLS has no explicit dynamics under $\m{H}$ and for most observables the two sectors of the Hilbert space corresponding to $\sigma^z=\pm$ decouple, their dynamics being described by the Hamiltonians $\m{H}[+]=\m{H}+V_{loc}=\m{H}_+$ and $\m{H}[-]=\m{H}$ respectively. However this is not the case for correlators of the TLS itself, such as for example 
 \be
\m{G}_{-+}(t',t)=\langle\Psi_{0\downarrow}\vert\, {\sigma}^-(t')\,\sigma^+(t)\vert\Psi_{0\downarrow}\rangle
\ee
 where the TLS operators are evolved with the full Hamiltonian $\m{H}[\sigma^z]$, i.e. 
 \be 
 \sigma^{+}(t)=e^{i\,\m{H}[\sigma^z]\,t}\,\,\sigma^{+}\,\,e^{-i\,\m{H}[\sigma^z]\,t}=
 e^{i\left(\m{H}+V_{\rm loc}\right)t}\, e^{-i\m{H}t}\,\left(1-\sigma^z\right)/2
 \ee
 while the average is taken over the state $ \vert\,\Psi_{0\downarrow}\rangle=\vert\Psi_0\rangle\otimes\vert \downarrow\rangle$. It is then easy to see that the TLS propagator (or core-hole Green's function in the fermionic language) takes the form of 
\be
\m{G}_{-+}(t',t)=\langle\Psi_0\vert\,e^{i\m{H}t'}\,e^{-i\left(\m{H}+V_{\rm loc}\right)\left(t'-t\right)}\,
e^{-i\m{H}t}\vert\Psi_0\rangle= \m{D}(t',t)
\ee

\section{Local Real-Time Green's Functions for a Quenched Luttinger Liquid}

In this section we compute the local real-time Green's functions (GF) of the Luttinger Liquid (LL) after a quench of $K,u$.
At time $t=0$ we have an equilibrium LL with the Hamiltonian
\be\label{eqn:H0}
H_{0}=\frac{u_0}{2\pi}\int\,dx\,\left[K_0\,\left(\pi\Pi\right)^2+\frac{1}{K_0}\,\left(\partial_x\phi\right)\right]
\ee
and for time $t>0$ we let the system evolve with the Hamiltonian
\be\label{eqn:H}
H= \frac{u}{2\pi}\int\,dx\,\left[K\,\left(\pi\Pi\right)^2+\frac{1}{K}\,\left(\partial_x\phi\right)\right]
\ee
To preserve Galilean invariance (which is not necessary for the formalism), we assume $u = v_F/K, u_0=v_F/K_0$.
The retarded (R), advanced (A) and Keldysh (K) component of the local Green's function at the impurity site $x=0$,  i.e. for the field $\phi(t)\equiv\phi(x=0,t)$ are defined as
\bea
G^R_{\phi\phi}(t,t')=-i\theta(t-t')\,\langle\,[\phi(t),\phi(t')\,]\rangle\\
G^A_{\phi\phi}(t,t')=i\theta(t'-t)\,\langle\,[\phi(t),\phi(t')\,]\rangle \\
G^K_{\phi\phi}(t,t')=-i\,\langle\,\left\{\phi(t),\phi(t')\,\right\}\rangle
\eea
where we notice that $G^A_{\phi\phi}(t,t')=G^R_{\phi\phi}(t',t)$. In these expressions, the time evolution of the field operator is done with respect to $H$ in Eq.~(\ref{eqn:H}) while the average is taken over
the ground state of $H_0$. We use a bosonization prescription where,
\begin{eqnarray}
&&\phi(x) =
-(N_{R}+N_{L})\frac{\pi x}{L}
-\frac{i\pi}{L}\sum_{p\neq0}\left(\frac{L|p|}{2\pi}\right)^{1/2}
\frac{1}{p}
e^{-\alpha|p|/2-ipx}\left(b_{p}^{\dagger} + b_{-p}\right), \\
&&\theta(x) =
(N_{R}-N_{L})\frac{\pi x}{L} +
\frac{i\pi}{L}\sum_{p\neq0}
\left(\frac{L|p|}{2\pi}\right)^{1/2}
\frac{1}{|p|}e^{-\alpha|p|/2-ipx}\left(b_{p}^{\dagger} - b_{-p}\right).
\end{eqnarray}
Let us denote a UV cutoff as $\Lambda = \frac{u}{\alpha}$.

In order to evaluate the real-time Green's function  of the field $\phi(x)$ it is useful to express the Heisenberg time evolution of the bosonic modes $b_{p},b^{\dagger}_{p}$
\be
b_p(t) =e^{i\,H\,t}\,b_p\,e^{-i\,H\,t}\,\qquad\,
b^{\dagger}_{-p}(t) =e^{i\,H\,t}\,b^{\dagger}_{-p}\,e^{-i\,H\,t}\,\qquad\,
\ee
in terms of the bosonic modes that diagonalize the initial Hamiltonian, $\eta_p,\eta^{\dagger}_{-p}$. After some simple algebra we get
\bea
\left(
\begin{array}{l}
b_p(t)\\
\\
b^{\dagger}_{-p}(t)
\end{array}\right)=
\left(
\begin{array}{lll}
f_p(t) & & -g_p(t) \\
 & &\\
-g_p^*(t) & & f^*_p(t)
\end{array} \right)
\left(
\begin{array}{l}
\eta_p\\
\\
\eta^{\dagger}_{-p}
\end{array}
\right)
\eea
where
\bea
f_p(t)= \cosh\beta_0\cos\left(u\vert p\vert\,t\right) - i \cosh\left(2\beta-\beta_0\right)\,\sin\left(u\vert p\vert\,t\right) \\
g_p(t)= \sinh\beta_0\cos\left(u\vert p\vert\,t\right) +i \sinh\left(2\beta-\beta_0\right)\,\sin\left(u\vert p\vert\,t\right)
\eea
with $e^{-2\beta_0}=K_0$ and $e^{-2\beta}=K$. Useful combinations of bosonic operators are
\bea
b^{\dagger}_p(t)-b_{-p}(t)= -A_p(t)\,\eta_{-p}+A^*_p(t)\,\eta^{\dagger}_p\\
b^{\dagger}_p(t)+b_{-p}(t)= B_p(t)\,\eta_{-p}+B^*_p(t)\,\eta^{\dagger}_p
\eea
where we have defined
\bea
A_p(t)= f_p(t)+g^{\star}_p(t)=\frac{1}{\sqrt{K_0}}\left(\cos\,u\vert p\vert\,t - i\frac{K_0}{K} \sin\,u\vert p\vert\,t  \right)\\
B_p(t)= f_p(t)-g^{\star}_p(t)=\sqrt{K_0}\left(\cos\,u\vert p\vert\,t - i\frac{K}{K_0} \sin\,u\vert p\vert\,t  \right)
\eea
In addition it is useful to remember that for a thermal occupation of the bosonic mode $\eta_{p}$ we have
\bea
\langle\,\eta_p\,\eta^{\dagger}_p\rangle_0 = \frac{1}{2}+\frac{1}{2}\,\coth\left(\frac{u\vert\,p\vert}{2T}\right)\\
\langle\,\eta^{\dagger}_p\,\eta_p\rangle_0 = -\frac{1}{2}+\frac{1}{2}\,\coth\left(\frac{u\vert\,p\vert}{2T}\right)\\
\eea

Inserting the definition of the field $\phi(x=0,t)$ in terms of the bosonic modes gives the following for the retarded Green's function
\be
G^R_{\phi\phi}(t>t')= -K\,\int_{0}^{\infty}\,\frac{dp}{\vert\,p\vert}\,e^{-\alpha\vert\,p\vert}\,
\sin u\,p\left(t-t'\right)
=-K\,\arctan\Lambda\left(t-t'\right)
\ee

We notice that $G^{R,A}$ does not depend on the initial condition and is time-translational-invariant (TTI) even for a non-equilibrium quench problem, an artifact of the non-interacting nature of the problem. In contrast  the Keldysh Green's function reads
\be
G^K_{\phi\phi}(t,t')=-\frac{i\,K_0}{2}\,
\int_{0}^{\infty}\,\frac{dp}{\vert\,p\vert}\,e^{-\alpha\vert\,p\vert}\,
\coth\left(\frac{u_0\vert\,p\vert}{2T}\right)\left[
\left(1-\frac{K^2}{K_0^2}\right)\,\cos up\left(t+t'\right)+
\left(1+\frac{K^2}{K_0^2}\right)\,\cos up\left(t-t'\right)
\right]
\ee
and depends as expected on the initial condition through $K_0$ and the
occupation probability of the boson modes in the initial state, here assumed to be
in equilibrium at a finite temperature $T$. $G^K$ shows a TTI component as well as a non-TTI one.

Similarly we can define the local Green's functions for the field $\theta(t)\equiv\theta(x=0,t)$
\bea
G^R_{\theta\theta}(t_1,t_2)=-i\theta(t_1-t_2)\langle\,[\theta(t_1),\theta(t_2)\,]\rangle\\
G^A_{\theta\theta}(t_1,t_2)=i\theta(t_2-t_1)\langle\,[\theta(t_1),\theta(t_2)\,]\rangle \\
G^K_{\theta\theta}(t_1,t_2)=-i\,\langle\,\left\{\theta(t_1),\theta(t_2)\,\right\}\rangle
\eea
One can immediately see that the above correlators for the $\theta$-field can be obtained from those for the $\phi$-field by making the substitution
\be
K_0\rightarrow\,\frac{1}{K_0}\,\qquad\,
K\rightarrow\,\frac{1}{K}\,\qquad\,
\ee
as one can confirm by a direct calculation.

Finally we will also be interested in computing the following correlator
\begin{eqnarray}
&&C_{ab}(t_1,t_2) = \langle e^{2i \phi_a(t_1)}e^{-2i \phi_b(t_2)}\rangle
\end{eqnarray}
where $a,b=\pm$. In general we can write it as
\begin{eqnarray}
&&C_{ab}(t_1,t_2) = e^{-2\left[\frac{iG^K(1,1)}{2} + \frac{iG^K(2,2)}{2}-i G^K(12)+ i a G^A(1,2)+i b G^R(1,2)\right]}
\end{eqnarray}
If we consider for example, $a=b=+$ and use the previous result we find
\begin{eqnarray}
C_{++}(1,2) = e^{-2\left[\frac{iG^K(1,1)}{2}+ \frac{iG^K(2,2)}{2}-iG^K(1,2)\right]}
e^{i 2K\,\left(sgn(t_1-t_2)
\tan^{-1}\Lambda(t_1-t_2)\right)}\\
\end{eqnarray}
We need the combination
\begin{eqnarray}
&&-2\left[\frac{iG^K(1,1)}{2}+ \frac{iG^K(2,2)}{2}-iG^K(1,2)\right]=\nonumber \\
&&= -2K_{neq}\,\ln{\sqrt{1+\Lambda^2(t_1-t_2)^2}}-K_{tr}
\ln\sqrt{\frac{(1+\Lambda^2(t_1+t_2)^2)^2}{(1+4\Lambda^2t_1^2)(1+4\Lambda^2t_2^2)}}
\end{eqnarray}
where we have introduced the coefficients
\begin{eqnarray}
K_{neq} = \frac{1}{2}K_0\left(1+\frac{K^2}{K_0^2}\right)\\
K_{tr} = \frac{1}{2}K_0\left(1-\frac{K^2}{K_0^2}\right)
\end{eqnarray}

\section{Forward Scattering Contribution to the OC Correlator: Unitary Transformation}

As anticipated in the main text, the evaluation of the transient OC correlator $\m{D}(t';t)$ greatly simplifies by noticing that the forward and backward scattering processes are decoupled. To see this we follow standard steps~\cite{VonDelftSchoeller_review98,GogolinNerseyanTsvelik_2004} and introduce first the even and odd combinations of the LL fields $\theta,\phi$, defined as
\bea
\phi_{\rm e/o}=\frac{\phi(x)\pm\phi(-x)}{\sqrt{2}}\\
\theta_{\rm e/o}= \frac{\theta(x)\pm\theta(-x)}{\sqrt{2}}
\eea
Then, it is convenient to introduce new chiral bosonic fields $
\Phi_{\rm s/a}(x)=\sqrt{K}\theta_{\rm o/e}(x)+\frac{1}{\sqrt{K}}\phi_{\rm e/o}(x)$
satisfying $\left[\Phi_{\rm s}(x),\Phi_{\rm s}(y)\right]=[\Phi_{\rm a}(x),\Phi_{\rm a}(y)]=-i\pi\,\mbox{sign}(x-y)\,$ and $
\left[\Phi_{\rm s}(x),\Phi_{\rm a}(y)\right]=0$. It is particularly convenient to write the local interaction in terms of these fields due to their well defined properties under inversion. In particular the local interaction may be written as
\be
V_{\rm loc}=g_{\rm fs}\,\sqrt{\frac{K}{2}}\,\partial_x\,\Phi_{\rm a}(x)\vert_{x=0}+
g_{\rm bs}\,\cos\,\sqrt{2K}\Phi_{\rm s}(x=0)\nonumber
\ee
While the bulk LL Hamiltonian becomes
\be
\m{H}=\frac{u}{4\pi} \int\,dx\,\sum_{\nu=s,a}\,\left(\partial_x\Phi_{\nu}\right)^2\equiv \sum_{\nu=s,a}\,\bar{\m{H}}_{\nu}
\ee
As a result, we find that the OC correlator factorizes into $\m{D}(t',t)=\m{D}_{\rm fs}(t',t)\,\m{D}_{\rm bs}(t',t)$.
We now analyze the forward scattering term for the case of a quenched non-equilibrium LL, which can be evaluated exactly
using the same trick introduced by Schotte and Schotte~\cite{SchotteSchottePR69} in their bosonization treatment of the Xray edge problem in Fermi Liquids. We aim to compute the correlator
\be
\m{D}_{\rm fs}(t',t)=
\langle\Psi_0\vert e^{i\,\m{\bar{H}}_at'}\,e^{-i\,\m{\bar{H}}_{a+}(t'-t)}e^{-i\,\m{\bar{H}}_at}\vert\Psi_0\rangle
\ee
where $\m{\bar{H}}_{\rm a+}$ contains the forward scattering potential term for the antisymmetric field $\Phi_{\rm a}$, i.e.
\be
\m{\bar{H}}_{\rm a+}=
\m{\bar{H}}_{\rm a}+V_{fs}=\frac{u}{4\pi}\int\,dx\,\left(\partial_x\Phi_{\rm a}\right)^2+g_{fs}\,\sqrt{\frac{K}{2}}\,\partial_x\,\Phi_{\rm a}(0)
\ee
The calculation can be done exactly by noticing that the forward scattering term can be eliminated by a unitary transformation $\Omega$ such that
\be\label{eqn:Omega}
\Omega^{\dagger}\,\bar{\m{H}}_{\rm a+}\,\Omega=\bar{\m{H}}_{\rm a}
\ee
Then one can write the above correlator as
\bea
\m{D}_{fs}(t',t)=\langle\Psi_0\vert\,e^{i\,\m{\bar{H}}_at'}\,\Omega\,\Omega^{\dagger}\,e^{-i\,\m{\bar{H}}_{a+}(t'-t)}\,
\Omega\,\Omega^{\dagger}\,e^{-i\,\m{\bar{H}}_at}\vert\Psi_0\rangle
=\nonumber\\=
\langle\Psi_0\vert\,e^{i\,\m{\bar{H}}_at'}\,\Omega\,e^{-i\,\m{\bar{H}}_{a}(t'-t)}\,
\Omega^{\dagger}\,e^{-i\,\m{\bar{H}}_at}\vert\Psi_0\rangle=
\langle\Psi_0\vert\,\Omega(t')\Omega^{\dagger}(t)\vert\Psi_0\rangle
\eea
The unitary operator $\Omega$ takes the form
\be
\Omega = \exp\left[i\lambda\Phi_a(0)\right]\,
\ee
with $\lambda=-\sqrt{\frac{K}{2}}\,g_{fs}/u$. This can be easily seen using the algebra of the fields $\Phi_{a}(x)$. In particular it is easy to see that $\Omega$ implements a shift
\bea
\Omega^{\dagger}\,\Phi_a(x)\,\Omega = \Phi_a(x)+\lambda\,\pi\,\mbox{sign}(x)\\
\Omega^{\dagger}\,\partial_x\Phi_a(x)\,\Omega = \partial_x\Phi_a(x)+2\lambda\,\pi\,\delta(x)
\eea
so that by plugging this result into Eq.~(\ref{eqn:Omega}) we see that by properly choosing $\lambda$ we can eliminate the forward scattering term.  Then, using the relation between the antisymmetric mode $\Phi_a(0)$ and the original degrees of freedom of the LL $\theta(0)$
\be
\Phi_a(x=0)=\sqrt{2K}\,\theta(x=0)
\ee
we finally obtain the correlator
\be
\m{D}_{fs}(t',t)=
\langle\Psi_0\vert\,e^{-i\eta\theta_+(0,t')}\,e^{i\eta\theta_+(0,t)}\vert\Psi_0\rangle=
\exp\left[-\frac{\eta^2}{2}\,\langle\,(\theta_+(t')-\theta_+(t))^2\rangle\right]
\,\qquad\,
 \eta=\frac{g_{fs}\,K}{u}
\ee
whose evaluation reduces to known correlation functions of the quenched LL which have been evaluated in the previous section.

\section{Backscattering contribution to the OC Correlator: Perturbative Calculation}

The backscattering contribution to the OC correlator reads $D_{bs}(t',t)=\exp\,C(t',t)$, with
\be
\m{C}(t',t)=\langle\Psi_0\vert\,T\,e^{-i\,\int_t^{t'}\,dt_1\,V_{\rm bs}(t_1)}\vert\Psi_0\rangle_{c}
\ee
This quantity admits a perturbative expansion, which can be written in terms of the basic correlators of the quenched bulk LL. We can evaluate the second order contribution with a calculation similar to the previous one
\be
\m{C}^{(2)}(t',t)=\frac{(-i)^2}{2!}\,g_{bs}^2\int_t^{t'}\,dt_1\,\int_t^{t'}\,dt_2
\langle\,\Phi_0\vert\,T\left[\cos\left(\gamma\phi(t_1)\right)\,\cos\left(\gamma\phi(t_2)\right)\right]\vert\Phi_0\rangle_{conn}
\ee
which can be written after some algebra using the results of previous sections as
\bea
\m{C}^{(2)}(t',t)=\frac{(-i)^2}{4}\,g_{bs}^2\int_t^{t'}\,dt_1\,\int_t^{t'}\,dt_2\,\,
\frac{\exp\left[2i\,K\,\mbox{sign}(t_1-t_2)\arctan\left(\Lambda(t_1-t_2)\right)\right]}
{\left(1+\Lambda^2\left(t_1-t_2\right)^2\right)^{K_{neq}}}
\left[\frac{\left(1+4\Lambda^2\,t_1^2\right)\left(1+4\Lambda^2\,t_2^2\right)}
{\left(1+\Lambda^2(t_1+t_2)^2\right)^2}\right]^{\frac{K_{tr}}{2}}
\eea
It is tempting to proceed as previously and shift the integration variables $t_{1,2}\rightarrow t_{1,2}-t$ to get
 (here we put $\tau=t'-t>0$)
\bea
&&\m{C}_t^{(2)}(\tau)=\nonumber \\
&&-\frac{g_{bs}^2}{4}\int_0^{\tau}\,dt_1\,\int_0^{\tau}\,dt_2\,\,
\frac{\exp\left[2i\,K\,\mbox{sign}(t_1-t_2)\arctan\left(\Lambda(t_1-t_2)\right)\right]}
{\left(1+\Lambda^2\left(t_1-t_2\right)^2\right)^{K_{neq}}}
\left[\frac{\left(1+4\Lambda^2\,(t_1+t)^2\right)\left(1+4\Lambda^2\,(t_2+t)^2\right)}
{\left(1+\Lambda^2(t_1+t_2+2t)^2\right)^2}\right]^{\frac{K_{tr}}{2}}
\eea
Let us look in more detail at this result. In order to proceed we do a change of variables in the integral
\bea
t_1=X+x\\
t_2=X-x
\eea
We notice that the first term is only a function of $t_1-t_2=2x$ (we set $\Lambda=1$ in the following)
\be
f(x)\equiv  \frac{\cos\left[2\,K\,\arctan\left(2x\right)\right]}
{\left(1+4x^2\right)^{K_{neq}}}
\ee
 while the second one depends on both $X,x$, and also on the absolute time $t$.
\be
G_t(x,X)=
\left[\frac{\left(1+\,4(X+x+t)^2\right)\left(1+4(X-x+t)^2\right)}
{\left(1+4(X+t)^2\right)^2}\right]^{\frac{K_{tr}}{2}}
\ee

\begin{figure}[t]
\begin{center}
\epsfig{figure=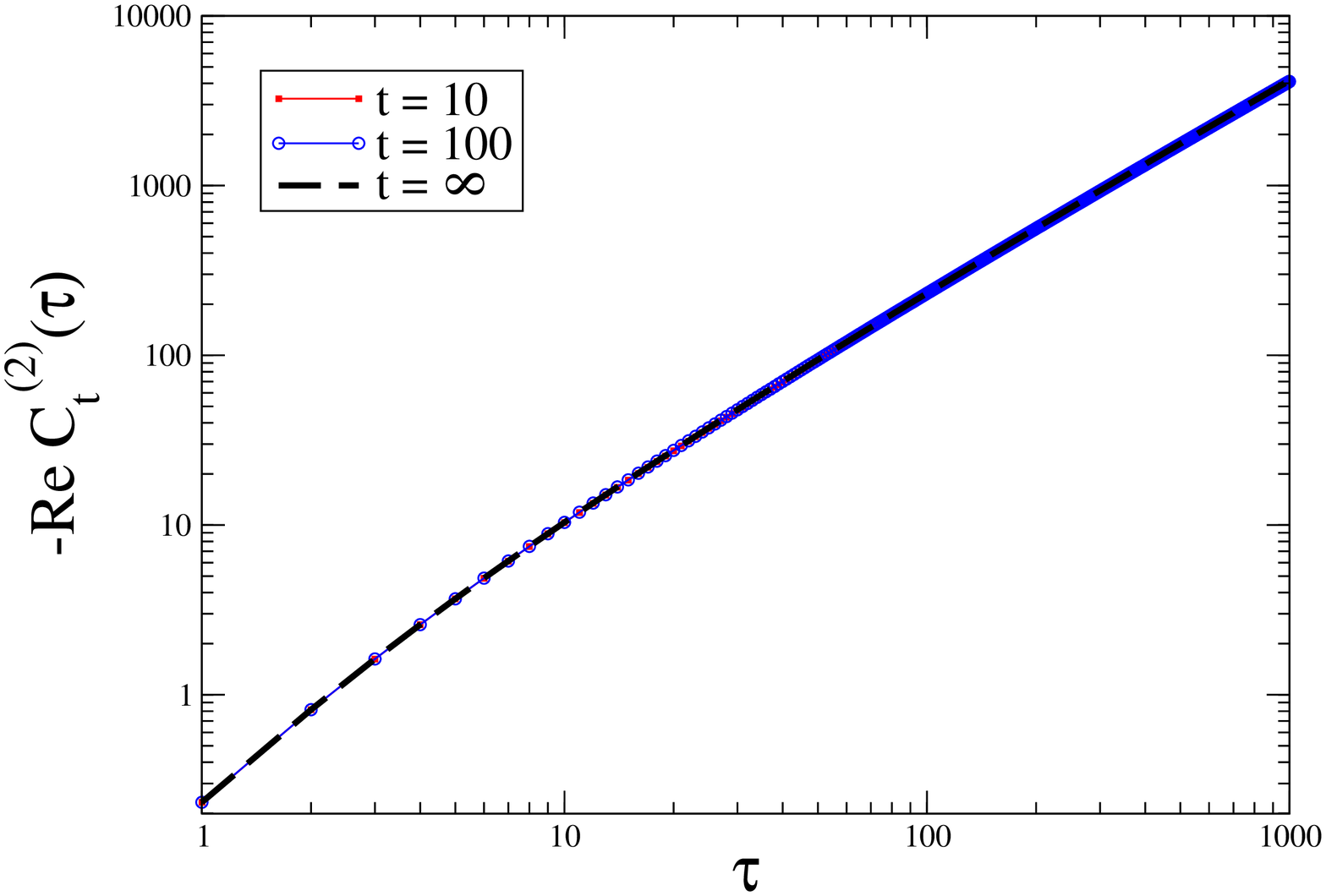,scale=0.3}
\epsfig{figure=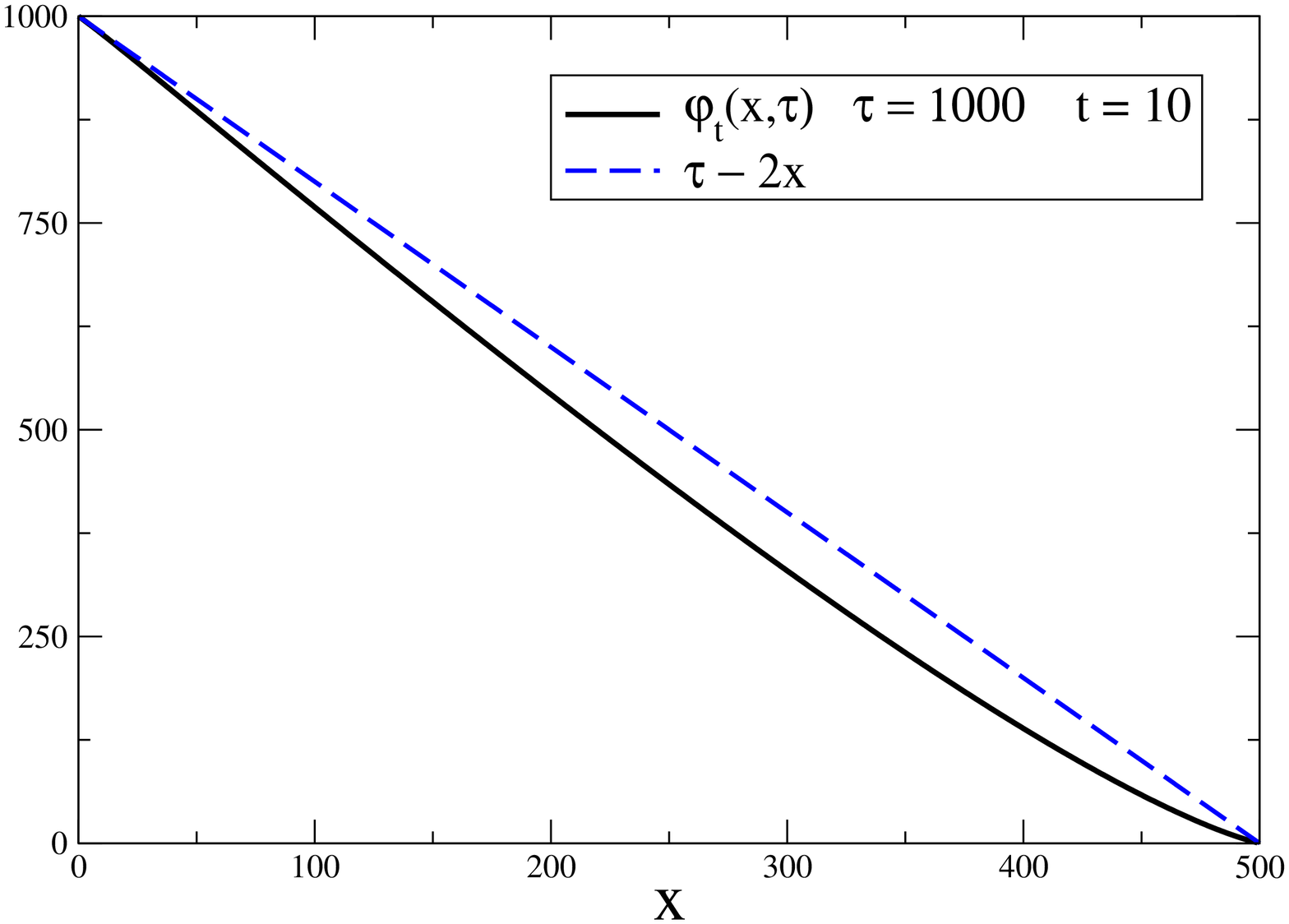,scale=0.3}
\caption{Left Panel: $C^{(2)}_t(\tau)$ correlator for $K_0=0.75, K=0.25$ and different values of the waiting time $t$. We see the transient factor in this case has no effect on the overall scaling with $\tau$, differently from the forward scattering case. In the left panel we show, for $K_0=0.75, K=0.25$ the function $\varphi_t(x,\tau)$ defined in the text and compare it to its long time limit $\varphi_{\infty}(x,\tau)$ to show that finite time corrections have little effect in the region of small $x$.}
\label{fig:fig1}
\end{center}
\end{figure}

 We then write the integral as
\bea
\m{C}_t^{(2)}(\tau)=-\frac{g^2_{bs}}{4}\,2\,
\left[\int_{-\tau/2}^{0}\,dx\,\int_{-x}^{x+\tau}
d\,X\,
f(x)\,
G_t(x,X)
+
\int_{0}^{\tau/2}\,dx\,\int_{x}^{\tau-x}
dX
f(x)\,
G_t(x,X)\right]=\\
\label{eqn:C2trans}
=-g^2_{bs}\,\int_{0}^{\tau/2}\,dx\,f(x)\,
\int_{x}^{\tau-x}
dX
G_t(x,X)\equiv
-g^2_{bs}\,\int_{0}^{\tau/2}\,dx\,f(x)\,\varphi_t(x,\tau)
\eea
where we defined
$$
\varphi_t(x,\tau)=\int_{x}^{\tau-x}
dX
G_t(x,X)
$$
and we also used the fact that
$$
G_t(x,X)=G_t(-x,X)\qquad
f(x)=f(-x)
$$
In the case of an infinite waiting time after the global quench, $t\rightarrow\infty$, the factor $G_t(x,X)$ approaches a constant and we get
$$
\varphi_{\infty}(x,\tau)=\int_{x}^{\tau-x}
dX=\tau-2x
$$
so that the evaluation of the correlator reduces to the single integral
\be
 \m{C}_{\infty}^{(2)}(\tau)=-g^2_{bs}\,\int_{0}^{\tau/2}\,dx\,\left(\tau-2x\right)\,f(x)\,
\ee
whose behavior has been discussed in the main text. In figure~\ref{fig:fig1} we evaluate numerically the correlator $C_t^{(2)}(\tau)$ and show that finite time corrections do not change the scaling in $\tau$ and the overall qualitative behavior of this correlator, quite differently from the forward scattering case. The reason is that, in the region where the integral~(\ref{eqn:C2trans}) takes its main contribution, i.e. for small $x$,  the function $\varphi_t(x,\tau)$ does not differ substantially from its infinite time limit $\varphi_{\infty}(x,\tau)$ (see figure~\ref{fig:fig1}, right panel).
\subsection{Equilibrium}

Let us discuss the equilibrium case, where analytical results can be obtained. Here $K_0=K$, which implies
$K_{neq}=K_{eq}$ and $K_{tr}=0$. In what follows we set $\gamma=2$ so that $K_{eq}=K$.
Then we can write, after some algebra (setting $\tau=t'-t$)
\bea
\m{C}^{(2)}(\tau)=\frac{(-i)^2}{4}\,g_{bs}^2\int_0^{\tau}\,dt_1\,\int_0^{\tau}\,dt_2\,\,
\frac{\exp\left[2i\,K\,\mbox{sign}(t_1-t_2)\arctan\left(\Lambda(t_1-t_2)\right)\right]}
{\left(1+\Lambda^2\left(t_1-t_2\right)^2\right)^{K}}
\\=-\frac{g_{bs}^2}{2}\,\int_0^{\tau}\,dx\,\left(\tau-x\right)
\frac{\exp\left[2i\,K\,\arctan\,x\right]}
{\left(1+x^2\right)^{K}}
\eea
The integral can be evaluated in closed form and we get for the real part
\be
\mbox{Re}\m{C}^{(2)}(\tau)=-\frac{g_{bs}^2}{2}\,
\frac{1}{2(K-1)\left(2K-1\right)}\left[
1+\frac{1}{\left(1+\tau^2\right)^{K}}\left(
(\tau^2-1)\cos\left(2K\arctan\tau\right)-2\tau\sin\left(2K\arctan\tau\right)
\right)
\right]
\ee
For large times $\tau\gg1$ we find
\be\label{eqn:C2_eq}
 \mbox{Re}{C}^{(2)}(\tau)\sim -\frac{g_{bs}^2}{2}\frac{1}{(2K-1)}\left[\frac{\cos(\pi\,K)\tau^{2(1-K)}+1-2\tau^{1-2K}\sin(\pi K)}{2(K-1)}\right]
\ee
This is an interesting result: we see that for $K>1$ the correlator goes to a constant at long times, i.e. the perturbation theory does not break down, which is consistent with the RG flow of the back-scattering being irrelevant. In contrast for $K<1$ the perturbative correction blows up at long times and the result~(\ref{eqn:C2_eq}) is only valid on short times scales while higher order corrections diverge even faster. One can extract a time scale $\tau_*$ which signals the breakdown of perturbation theory and the flow to strong coupling, given by the condition
\be
\m{C}^{(2)}(\tau_*)\sim 1\qquad\longrightarrow\qquad \tau_*\sim \frac{1}{\Lambda}\left(\frac{\Lambda}{g_{bs}}\right)^{1/(1-K)}
\ee
which is consistent with the scale one obtains from the RG (see below). The situation at $K=1$ is of course peculiar, since at the critical point the RG flow is marginal and we should expect a logarithmic behavior. The result~(\ref{eqn:C2_eq}) for $K\rightarrow1$ confirms this expectation
\be
 \m{C}^{(2)}(\tau)\sim -g_{bs}^2\log\tau
\ee
which is indeed consistent with the well known properties of the $\m{C}$ correlator for a non-interacting ($K=1$) Fermi gas which is log singular to all orders in perturbation theory. A logarithmic long time behavior of $\m{C}(\tau)$ immediately gives a power law behavior for the core-hole propagator $\m{D}(\tau)$, consistent with the Nozieres and De Dominicis solution of the X-ray edge problem.

\subsubsection{Equilibrium Finite T}

Before concluding, let us evaluate how the correlator $C^{(2)}(\tau)$ behaves in equilibrium at finite temperature $T$.
For this we only need the Keldysh component of the local field Green's function which now reads
\be
G^K_{\phi\phi}(t_1-t_2)= -iK_0\int_0^{\infty}\frac{dp}{p}\,e^{-\alpha\,p}\coth\left(\frac{\eps_p}{2T}\right)\,\cos\eps_p(t_1-t_2)
\ee
Using this result and repeating the previous steps we get
\be
\m{C}^{(2)}(\tau)=\frac{(-i)^2}{4}\,g^2_{bs}\int_0^{\tau}\,dt_1\,\int_0^{\tau}\,dt_2\,\,
\exp\left[2i\,K\,\mbox{sign}(t_1-t_2)\arctan\left(t_1-t_2\right)\right]
\exp\left(-2K\,f_T\left(t_1-t_2\right)\right)
\ee
where
$$
f_T(x)=\frac{1}{2}\log\left(1+x^2\right)+2\log\Gamma\left(1+T\right)-
\log\left(\Gamma\left(1+T(1-ix)\right)-\Gamma\left(1+T(1+ix)\right)\right)
$$
\begin{figure}[t]
\begin{center}
\epsfig{figure=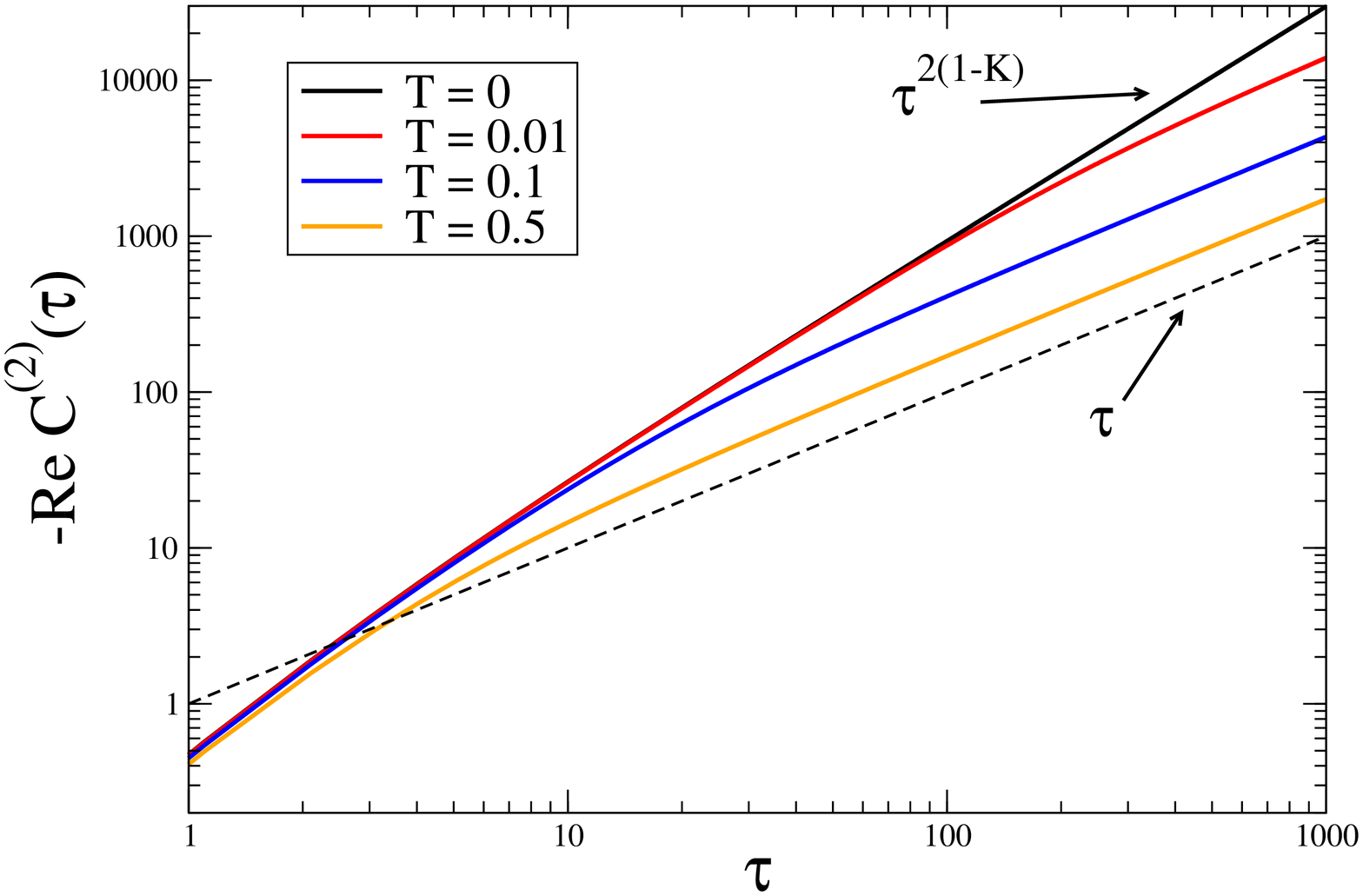,scale=0.3}
\epsfig{figure=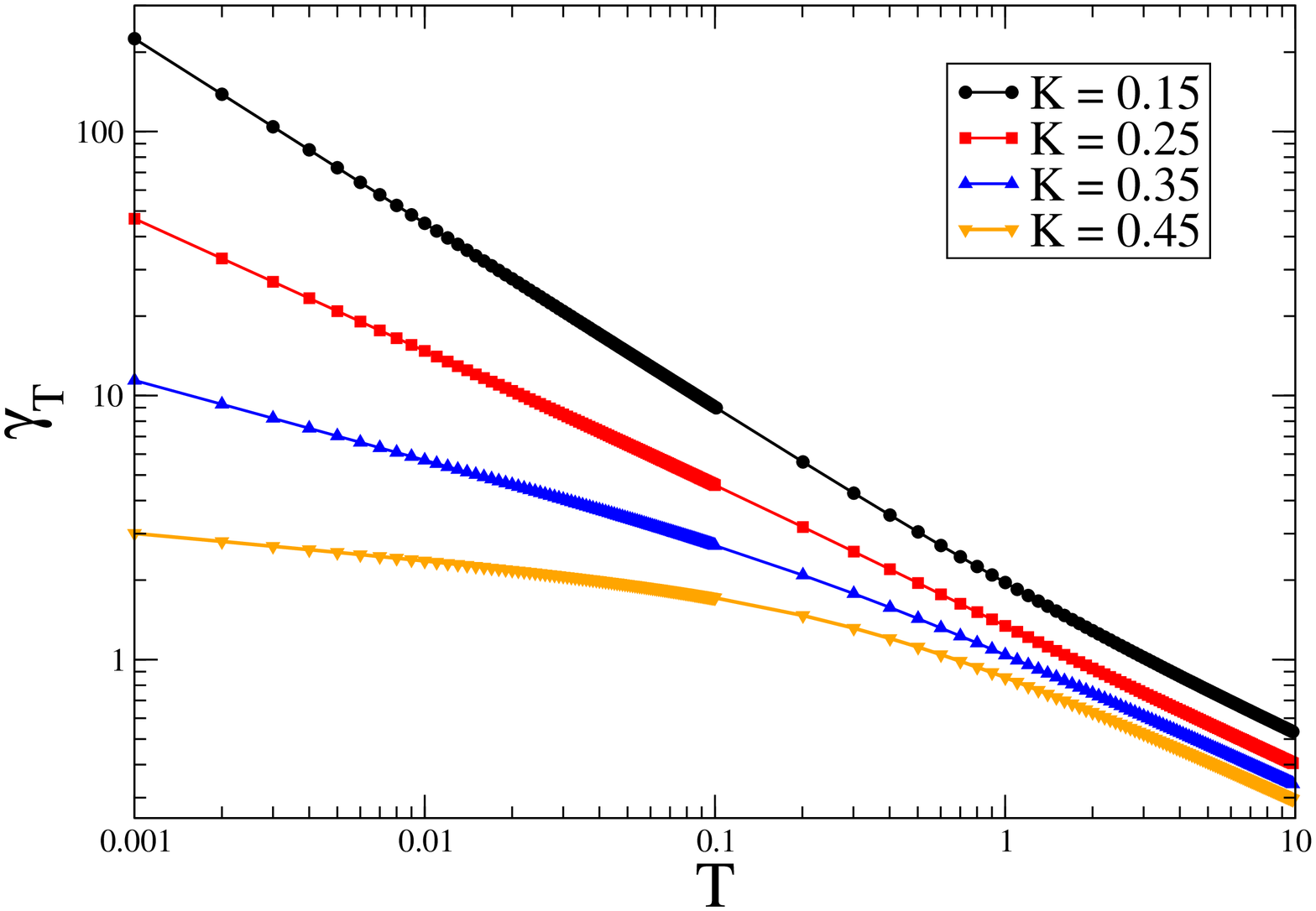,scale=0.3}
\caption{Left Panel: $C^{(2)}(\tau)$ correlator in equilibrium at finite temperature for $K=0.25$. We notice the crossover at long times to a linear behaviour, $C^{(2)}(\tau)\sim-\gamma_T\tau$. Right Panel: Finite temperature relaxation rate $\gamma_T$ as a function of temperature and for different values of $K$. In the relevant phase, $K<1/2$, the relaxation diverge at low temperature as a power law $\gamma_T\sim T^{2K-1}$.}
\label{fig:fig2}
\end{center}
\end{figure}
After some simple manipulations we obtain the following for the real part
\be
\mbox{Re}\,\m{C}^{(2)}(\tau)=-\frac{g^2_{bs}}{2}\int_0^{\tau}\,dy\left(\tau-y\right)
\frac{\cos\left[2\,K\,\arctan y\right]}{(1+y^2)^{K}}
\left[\frac{\Gamma(1+T(1-iy))\Gamma(1+T(1+iy))}{\Gamma^2(1+T)}
\right]^{2K}
\ee
We could not find a closed analytical form for this integral valid for any temperature and time, however from the numerical evaluation (see figure~\ref{fig:fig2}, left panel) we find that, at any finite $T$, the leading long-time behavior is linear for any $K$
\be
\mbox{Re}\,\m{C}^{(2)}(\tau\gg\tau_{th})\sim -\gamma_T\,\tau
\ee
The dynamics of $\mbox{Re}\,C^{(2)}(\tau)$ is interesting, with a clear time scale $\tau_{th}(T)$ controlling the crossover from the intermediate time "zero temperature" regime to the "finite temperature" long time regime, a time scale which diverges as $T\rightarrow0$. The slope of the linear term can be seen as a relaxation rate $\gamma_T$ which can be written as
\be
\gamma_T=\frac{g^2_{bs}}{2}\int_0^{\infty}\,dy \frac{\cos\left[2\,K\,\arctan y\right]}{(1+y^2)^{K}}
\left[\frac{\Gamma(1+T(1-iy))\Gamma(1+T(1+iy))}{\Gamma^2(1+T)}
\right]^{2K}
 \ee
 Numerically (see figure~\ref{fig:fig2}, right panel) we find a low temperature behavior for $\gamma_T$ consistent with a power law, i.e.
 $\gamma_T\sim g_{bs}^2 T^{2K-1}$. This result is also consistent with the fact that the renormalization group
 flows in equilibrium are cut-off by the temperature. Thus, at finite temperature, one may replace the bare
 couplings by the renormalized coupling $g_{bs}T^{K-1}$.

\section{Time Dependent RG for the backscattering potential}

For the RG formulation it is convenient to use a Keldysh action formalism.  To this extent we write the Keldysh generating functional for the problem
\be
\m{Z}=\int\,D\phi(x,t)\,\exp\left(i\m{S}[\phi(x,t)]\right)
\ee
where the action $\m{S}$ is defined on the two branches of the Keldysh contour $\m{C}$, and the time $t$ is a contour time living on $\m{C}$ (we will move to the RAK basis later). The action reads
\be
\m{S}=\frac{1}{2\pi}\int\,dx\,\int_{\m{C}}\,\,\frac{dt}{K}\,\left[\frac{1}{u}\left(\partial_t\phi\right)^2-u\,\left(\partial_x\phi\right)^2\right]+\frac{g_{bs}\,u}{\alpha}\,\int_{\m{C}}\,dt\,\cos\left(2\phi(0,t)\right)
\ee
 Since the non-linearity is only acting at a single point $x=0$ it is useful to integrate out all the bulk modes of the field, $\phi(x\neq0,t)$ and obtain an effective action for the boundary field $\phi(t)\equiv\phi(x=0,t)$,
\be
\m{Z}=\int\,D\phi(t)\,\exp\left(i\m{S}_{eff}[\phi(t)]\right)
\ee
In the next section we derive the effective Keldysh action for this impurity problem.

\subsection{Effective Action for the Local Impurity Degree of Freedom}

In this section we derive the effective action for the local impurity by integrating out all the bulk modes other than the field at $x=0$. To this extent we follow Ref.~\onlinecite{GogolinNerseyanTsvelik_2004} and introduce a delta function constraint
\be
\m{Z}=\int\,D\phi(x,t)\,\exp\left(i\m{S}[\phi(x,t)]\right)\,\int\,D\phi(t)\,
\delta\left[\phi(x=0,t)-\phi(t)\right]
\ee
Using an integral representation of the constraint
\be
 \delta\left[\phi(x=0,t)-\phi(t)\right]=\int\,D\,\lambda(t)\,\exp\left[i\int_{\m{C}}\,dt\,\lambda(t)\left(\phi(x=0,t)-\phi(t)\right)\right]
\ee
we can write the partition function as
\be
 \m{Z}=\int\,D\phi(t)\,\exp\left(i\m{S}_{bs}[\phi(t)]\right)\,\int\,D\,\lambda(t)\,
 \exp\left[-i\int_{\m{C}}\,dt\,\lambda(t)\,\phi(t)\right]\,\Gamma[\lambda(t)]
\ee
where $\m{S}_{bs}$ is the backscattering action and $\Gamma[\lambda(t)]$ reads
\be
 \Gamma[\lambda(t)]=\int\,D\phi(x,t)\,\exp\left[i\m{S}_{LL}[\phi(x,t)]+i\,\int\,dx\,\int_{\m{C}}\,dt\,\lambda(x,t)\,\phi(x,t)\right]
\ee
which is nothing but the generating functional of a Luttinger Liquid in a source $\lambda(x,t)=\delta(x)\,\lambda(t)$ that couples to the field at $x=0$. Since the theory is gaussian the generating functional can be easily obtained and reads
\be\label{eqn:Gamma}
\Gamma[\lambda(t)]=
\exp\left(-\frac{i}{2}\int_{\m{C}}\,dt_1\,\int_{\m{C}}\,dt_2\,
\lambda(t_1)\,G(x=y=0;t_1,t_2)\,\lambda(t_2)
\right)
\ee
where $G(x-y;t,t;')$ is the contour ordered Green's function of the non-equilibrium LL
\be
 G(x-y;t,t')=-i\langle\,T_{\m{C}}\phi(x,t)\,\phi(x,t')\rangle
\ee
 Finally, using this result we can perform the gaussian integral in Eq.~(\ref{eqn:Gamma}) over the auxiliary field $\lambda(t)$ to get
\be
\int\,D\,\lambda(t)\,
 \exp\left[-i\int_{\m{C}}\,dt\,\lambda(t)\,\phi(t)\right]\,\Gamma[\lambda(t)]=
 \exp\left(\frac{i}{2}\int_{\m{C}}\,dt_1\,\int_{\m{C}}\,dt_2\,
\phi(t_1)\,G^{-1}(x=y=0;t_1,t_2)\,\phi(t_2)
\right)
\ee
which gives the effective action
\be
\m{S}_{eff}= \frac{1}{2}\int_{\m{C}}\,dt\,\int_{\m{C}}\,dt'\,\phi(t)\,G^{-1}(x=y=0;t,t')\,\phi(t')+
\frac{g_{bs}\,u}{\alpha}\,\int_{\m{C}}\,dt\,\cos\left(2\,\phi(t)\right)
\ee

\subsection{Derivation of RG equations from the action}

Here we sketch the derivation of the RG equations from the effective action. We do the computation for a generic non-linear term
\be
S_{bs}=\frac{g_{bs}\,u}{\alpha}\,\int_{\m{C}}\,dt\,\cos\left(\gamma\,\phi(t)\right)
\ee
where the physical situation we are interested in corresponds to $\gamma=2$.

The quadratic part of the action may be schematically written as
\begin{eqnarray}
\int_0^t dt_1 \delta(t_1) \phi f(\Lambda t_1)\phi
\end{eqnarray}
where the $\phi$ fields are dimensionless scalar fields.

We split the fields into slow ($\phi^<$) and fast ($\phi^{>}$) components
\begin{eqnarray}
\phi_{\pm} = \phi^{<}_{\pm} + \phi^{>}_{\pm}
\end{eqnarray}
where the fast correlators are related in a simple way to derivatives of the full correlators,
\begin{eqnarray}
G^> = d\Lambda \frac{d G}{d\Lambda}
\end{eqnarray}

Then the action becomes
\begin{eqnarray}
\int_0^t dt_1 \delta(t_1) \phi^< f\left[(\Lambda -d\Lambda) t_1\right]\phi^< + {\rm fast-fields}\\
= \int_0^t dt_1 \delta(t_1) \phi^< f\left[\frac{(\Lambda -d\Lambda)}{\Lambda} \Lambda t_1\right]\phi^< + {\rm fast-fields}
\end{eqnarray}
Now define $t_1^{\prime}= t_1\left(\frac{\Lambda-d\Lambda}{\Lambda}\right)$, then the action becomes
\begin{eqnarray}
\int_0^{t\frac{\Lambda-d\Lambda}{\Lambda}} dt_1^{\prime} \delta(t_1^{\prime})
\phi^< f\left[\Lambda t_1^{\prime}\right]\phi^< + {\rm fast-fields}
\end{eqnarray}
Now let us make the arguments dimensionless $\bar{t}_1 = t_1^{\prime}\Lambda$. Then the action becomes
\begin{eqnarray}
\int_0^{t(\Lambda-d\Lambda)} d\bar{t_1} \delta(\bar{t}_1)
\phi^< f\left[\bar{t}_1\right]\phi^< + {\rm fast-fields}
\end{eqnarray}
Thus the quadratic part of the action goes back to itself provided that the time is rescaled according to
its engineering dimensions $t \Lambda \rightarrow t (\Lambda - d\Lambda)$.

In the presence of non-linearities, the fast fields may be integrated out perturbatively. Here we carry
this out to ${\cal O}(g^2_{bs})$
to obtain
\begin{eqnarray}
S = S_0^{<} + \delta S^{<} + S^<_{bs}
\end{eqnarray}
where
\begin{eqnarray}
&&S_0^{<} = \frac{1}{2}\int_0^{t} dT_m\left[
- 2\eta \phi_q^<\partial_{T_m}\phi_{cl}^< +i 4 \eta T_{eff}\left(\phi_q^<\right)^2\right]
\end{eqnarray}
with $\eta=\frac{2}{\pi K}, T_{eff}=0$ initially.
The back-scattering in terms of the slow fields is
\begin{eqnarray}
S_{bs}^<=g_{bs}\Lambda\int_0^t dt_1
\left[\cos\gamma\phi_-^<(1) - \cos\gamma\phi_+^<(1)\right]
e^{-\frac{\gamma^2}{4}\langle\left(\phi_{cl}^>(1)\right)^2\rangle}
\end{eqnarray}

The correction to the quadratic action from integrating the fast fields is,
\begin{eqnarray}
&&\delta S^<=+i\frac{g^2_{bs}\Lambda^2}{2}\int_0^t dt_1\int_0^t dt_2
\theta(t_1-t_2)
:\cos\left(\gamma\phi_-^<(1) -\gamma\phi_-^{<}(2)\right):
e^{-\frac{\gamma^2}{2}\langle(\phi_-(1)-\phi_-(2))^2\rangle}\nonumber \\
&&\times \left[1-e^{-\gamma^2 \langle \phi_-^>(1)\phi_-^>(2)\rangle}\right]\\
&&+ \frac{ig^2_{bs}\Lambda^2}{2}\int_0^t dt_1\int_0^t dt_2
\theta(t_2-t_1)
:\cos\left(\gamma\phi_+^<(1) -\gamma\phi_+^{<}(2)\right):
e^{-\frac{\gamma^2}{2}\langle(\phi_+(1)-\phi_+(2))^2\rangle}\nonumber \\
&&\times \left[1-e^{-\gamma^2 \langle \phi_+^>(1)\phi_+^>(2)\rangle}\right]\\
&&-\frac{ig^2_{bs}\Lambda^2}{2}\int_0^t dt_1\int_0^t dt_2
\{\theta(t_1-t_2)+ \theta(t_2-t_1)\}
:\cos\left(\gamma\phi_+^<(1) -\gamma\phi_-^{<}(2)\right):
e^{-\frac{\gamma^2}{2}\langle(\phi_+(1)-\phi_-(2))^2\rangle}\nonumber \\
&&\times \left[1-e^{-\gamma^2 \langle \phi_+^>(1)\phi_-^>(2)\rangle}\right]
\end{eqnarray}

Explicit expressions for the fast correlators are
\begin{eqnarray}
&&\langle\left(\phi_{cl}^>(t)\right)^2 \rangle = \frac{d\Lambda}{\Lambda}
\left[\frac{K_0}{2}\left(1+\frac{K^2}{K_0^2}\right)+\frac{K_0}{2}\left(1-\frac{K^2}{K_0^2}\right)
\biggl\{\frac{1}{1+\left(2t\Lambda\right)^2}\biggr\}\right]\\
&&\xrightarrow{t\ll 1/\Lambda } \frac{d\Lambda}{\Lambda}K_0\\
&&\xrightarrow{t\gg 1/\Lambda} \frac{d\Lambda}{\Lambda}
\frac{K_0}{2}\left(1+\frac{K^2}{K_0^2}\right)
\end{eqnarray}
For the non-local fast correlators, we have
\begin{eqnarray}
&&\langle \phi_{cl}^>(1) \phi_{cl}^>(2)\rangle = \frac{d\Lambda}{\Lambda}
\left[\frac{K_0}{2}\left(1+\frac{K^2}{K_0^2}\right)\frac{1}
{1 + \Lambda ^2(t_1-t_2)^2} +
\frac{K_0}{2}\left(1-\frac{K^2}{K_0^2}\right)\frac{1}
{1 + \Lambda^2(t_1+t_2)^2} \right]\\
&&\langle \phi_{cl}^>(1) \phi_{q}^>(2)\rangle = -i{K}\frac{d\Lambda}{\Lambda}\theta(t_1-t_2)\frac{\Lambda(t_1-t_2)}
{1 + \Lambda^2(t_1-t_2)^2}
\\
&& \langle \phi_{q}^>(1) \phi_{cl}^>(2)\rangle = i{K}\frac{d\Lambda}{\Lambda}\theta(t_2-t_1)
\frac{\Lambda(t_1-t_2)}
{1 + \Lambda^2(t_1-t_2)^2}
\end{eqnarray}

In the next step we define new variables
$T_m= \frac{t_1+t_2}{2}, \tau = t_1-t_2$
and perform a gradient expansion in $T_m$ to obtain,
\begin{eqnarray}
\delta S^{<} = \delta S^<_{0}+\delta S_{T_{eff}}^{<} + \delta S_{\eta}^{<}
\end{eqnarray}
where
\begin{eqnarray}
&&\delta S_{0}^{<} = \frac{g^2_{bs}\Lambda^2\gamma^2}{2}\frac{d\Lambda}{\Lambda}
\left[\int_0^{t/2} dT_m \int_{-2T_m}^{2T_m} d\tau + \int_{t/2}^tdT_m\int_{-2(t-T_m)}^{2(t-T_m)}d\tau \right]\nonumber \\
&&\times \theta(\tau)\left[\left(\tau\partial_{T_m} \phi_{cl}^<\right) \left(\tau\partial_{T_m} \phi_{q}^<\right)\right]
\times Im\left[e^{-\frac{\gamma^2}{2}\langle \left[\phi_+(0,T_m+\tau/2)-\phi_-(0,T_m-\tau/2)\right]^2\rangle}
F(T_m,\tau)\right]
\end{eqnarray}
$\delta S_0^<$ corresponds to generation of irrelevant terms of the form $\phi_{q}\partial_{T_m}^2 \phi_{cl}$. While
\begin{eqnarray}
&& \delta S_{T_{eff}}^{<}= \frac{ig^2_{bs}\Lambda^2\gamma^2}{2}\frac{d\Lambda}{\Lambda}
\left[\int_0^{t/2} dT_m \int_{-2T_m}^{2T_m} d\tau + \int_{t/2}^tdT_m\int_{-2(t-T_m)}^{2(t-T_m)}d\tau \right]\nonumber \\
&&\times \left(\phi_q^{<}(0,T_m)\right)^2Re\left[e^{-\frac{\gamma^2}{2}\langle \left[\phi_+(0,T_m+\tau/2)-\phi_-(0,T_m-\tau/2)\right]^2\rangle}
F(T_m,\tau)\right]
\end{eqnarray}
which corresponds to generation of a term of type $\phi_q^2$ whose physical meaning is noise. Finally,
\begin{eqnarray}
&&\delta S_{\eta}^{<} = \frac{g^2_{bs}\Lambda^2\gamma^2}{2}\frac{d\Lambda}{\Lambda}
\left[\int_0^{t/2} dT_m \int_{-2T_m}^{2T_m} d\tau + \int_{t/2}^tdT_m\int_{-2(t-T_m)}^{2(t-T_m)}d\tau \right]\nonumber \\
&&\times \phi_q^<(0,T_m)\tau \partial_{T_m}\phi_{cl}^<(0,T_m)
Im\left[e^{-\frac{\gamma^2}{2}\langle \left[\phi_+(0,T_m+\tau/2)-\phi_-(0,T_m-\tau/2)\right]^2\rangle}
F(T_m,\tau)\right]
\end{eqnarray}
which gives a correction to the local dissipation.
Note that,
\begin{eqnarray}
&&F(T_m,\tau) = 2K_{neq}\left(\frac{1}{1 + \Lambda^2\tau^2}\right)
+ 2K_{tr}\left(\frac{1}{1 + 4 \Lambda^2T_m^2}\right) -2 i K_{eq}\left(\frac{\Lambda\tau}{1 + \Lambda^2 \tau^2}\right)
\end{eqnarray}
whereas
\begin{eqnarray}
&&e^{-\frac{\gamma^2}{2}\langle \left[\phi_+(0,T_m+\tau/2)-\phi_-(0,T_m-\tau/2)\right]^2\rangle}
= \left[\frac{1}{{1 + \Lambda^2 \tau^2}}\right]^{K_{neq}}
\nonumber \\
&& \times \left[\frac{\sqrt{1 + \Lambda^2(2T_m+\tau)^2}}{\sqrt{1 + \Lambda^2 (2T_m)^2}}
\frac{\sqrt{1 + \Lambda^2(2T_m-\tau)^2}}{\sqrt{1 + \Lambda^2 (2T_m)^2}}\right]^{K_{tr}}e^{-2i K_{eq}\tan^{-1}\left(\Lambda\tau\right)}
\end{eqnarray}
where $Re[A] = (A+A^*)/2, Im[A] = (A-A^*)/(2i)$.

In order to arrive at a physical interpretation of dissipation and noise, it suffices to work with
the first half of the box $\int_0^{t/2}dT_m$. At long enough times, the coefficients become time-translationally invariant,
and therefore identical in the two parts of the box.

Thus, collecting all terms we find,
\begin{eqnarray}
&&S^<_{bs} = g_{bs}\Lambda \int_0^{t} dT_m
\left[\cos\gamma\phi_-^<(T_m) - \cos\gamma\phi_+^<(T_m)\right]
e^{-\frac{\gamma^2}{4}\langle\left(\phi_{cl}^>(T_m)\right)^2\rangle}\\
&&\delta S^<_0 = \frac{g^2_{bs}\gamma^2}{2\Lambda}\frac{d\Lambda}{\Lambda}
\int_0^{t} dT_m
\left[
- I_{T_m}(T_m)
\left(\partial_{T_m} \phi_{cl}^<\right) \left(\partial_{T_m} \phi_{q}^<\right)\right]\\
&& \delta S_{T_{eff}}^{<}= \frac{ig^2_{bs}\gamma^2\Lambda}{2}\frac{d\Lambda}{\Lambda}
\int_0^{t} dT_m
\left(\phi_q^{<}\right)^2 I_{T_{eff}}(T_m)\\
&& \delta S_{\eta}^{<} = -\frac{g^2_{bs}\gamma^2}{2}\frac{d\Lambda}{\Lambda}
\int_0^{t} dT_m
\phi_q^<\left[\partial_{T_m}\phi_{cl}^<\right] I_{\eta}(T_m)
\end{eqnarray}
where
\begin{eqnarray}
&&I_{T_m}(T_m) = -\Lambda^3\int_{0}^{2T_m} d\tau \tau^2
Im\left[e^{-\frac{\gamma^2}{2}\langle \left[\phi_+(0,T_m+\tau/2)-\phi_-(0,T_m-\tau/2)\right]^2\rangle}
F(T_m,\tau)\right]\\
&&I_{T_{eff}}(T_m)= \Lambda \int_{-2T_m}^{2T_m} d\tau
Re\left[e^{-\frac{\gamma^2}{2}\langle \left[\phi_+(0,T_m+\tau/2)-\phi_-(0,T_m-\tau/2)\right]^2\rangle}
F(T_m,\tau)\right]\\
&&I_{\eta}(T_m)= -\Lambda^2\int_{-2T_m}^{2T_m} d\tau  \tau
Im\left[e^{-\frac{\gamma^2}{2}\langle \left[\phi_+(0,T_m+\tau/2)-\phi_-(0,T_m-\tau/2)\right]^2\rangle}
F(T_m,\tau)\right]
\end{eqnarray}

At the next step we rescale the cut-off back to the original value of $\Lambda$, and in the
process rescale time $T_m \rightarrow \frac{\Lambda}{\Lambda^{\prime}}(T_m)$.
In other words we may write $f(\Lambda^{\prime}t) = f(\Lambda t\frac{\Lambda^{\prime}}{\Lambda})$.
Next on rescaling $t\rightarrow t \frac{\Lambda}{\Lambda^{\prime}}$
we obtain $f(\Lambda t)$. This recaling affects the upper-limit of the integration in time which goes from $T_m\Lambda \rightarrow
T_m \Lambda^{\prime}$.
Note that this rescaling is not necessary in expressions for $\delta S^<_{0,T_{eff},\eta}$ as they are already
of ${\cal O}\left(\frac{d\Lambda}{\Lambda}\right)$. We also express everything in dimensionless units of
$\bar{T} = T \Lambda$.
Thus to summarize one obtains,
\begin{eqnarray}
S_0^{<} &&=  \frac{1}{2}\int_0^{t\Lambda \left(\frac{\Lambda^{\prime}}
{\Lambda}\right)} d
\bar{T}_m\left[- 2\eta\phi_q^<\partial_{\bar{T}_m}\phi_{cl}^<
+i 4 \eta T_{eff}\left(\frac{\Lambda}{\Lambda^{\prime}}\right)
\left(\phi_q^<\right)^2\right]
\end{eqnarray}
and
\begin{eqnarray}
&&S^<_{bs} = g_{bs} \left(\frac{\Lambda}{\Lambda^{\prime}}\right)
\int_0^{t\Lambda \left(\frac{\Lambda^{\prime}}{\Lambda}\right)} d\bar{T}_m
\left[\cos\gamma\phi_-^<(T_m) - \cos\gamma\phi_+^<(T_m)\right]
e^{-\frac{\gamma^2}{4}\langle\left(\phi_{cl}^>(T_m)\right)^2\rangle}\\
&&\delta S^<_0 = \frac{g^2_{bs}\gamma^2}{2}\frac{d\Lambda}{\Lambda}
\int_0^{t\Lambda} d\bar{T}_m
\left[ - I_{T_m}(T_m)
\left(\partial_{\bar{T}_m} \phi_{cl}^<\right) \left(\partial_{\bar{T}_m} \phi_{q}^<\right)\right]\\
&& \delta S_{T_{eff}}^{<}= \frac{ig^2_{bs}\gamma^2}{2}\frac{d\Lambda}{\Lambda}
\int_0^{t\Lambda} d\bar{T}_m
\left(\phi_q^{<}\right)^2 I_{T_{eff}}(T_m)\\
&& \delta S_{\eta}^{<} = -\frac{g^2_{bs}\gamma^2}{2}\frac{d\Lambda}{\Lambda}
\int_0^{t\Lambda} d\bar{T}_m
\phi_q^<\left[\partial_{\bar{T}_m}\phi_{cl}^<\right] I_{\eta}(T_m)
\end{eqnarray}
Since,
\begin{eqnarray}
\frac{\Lambda}{\Lambda^{\prime}} = e^{d\ln(l)}\\
\frac{d\Lambda}{\Lambda} = \frac{\Lambda-\Lambda^{\prime}}{\Lambda}=d\ln(l)
\end{eqnarray}
Therefore the RG equations in terms of dimensionless variables such as
$T_m \rightarrow T_m \Lambda, \eta \rightarrow \eta , T_{eff}\rightarrow T_{eff}/\Lambda$
are
\begin{eqnarray}
\frac{dg_{bs}}{d\ln{l}} = g_{bs}\left[1-\left(K_{neq} + \frac{K_{tr}}{1+4 T_m^2}\right)\right]
\label{rg1}\\
\frac{d\eta}{d\ln{l}} = \frac{g^2_{bs}\gamma^2}{2}I_{\eta}(T_m)\label{rg2}\\
\frac{d(\eta T_{eff})}{d\ln{l}} = \eta T_{eff} + \frac{g^2_{bs}\gamma^2}{4}I_{T_{eff}}(T_m)
\label{rg3}\\
\frac{dT_m}{d\ln{l}} = -T_m \label{rg4}
\end{eqnarray}
Rewriting $\eta$ in terms of $K$ by noting that $\eta = \frac{2}{\pi K}$, the RG equations become,
\begin{eqnarray}
\frac{dg_{bs}}{d\ln{l}} = g_{bs}\left[1-\left(K_{neq} + \frac{K_{tr}}{1+4 T_m^2}\right)\right]
\label{rg1a}\\
\frac{d\left(1/K\right)}{d\ln{l}} = \frac{\pi g^2_{bs}\gamma^2}{4}I_{\eta}(T_m)\label{rg2a}\\
\frac{dT_{eff}}{d\ln{l}} = T_{eff} + \frac{\pi g^2_{bs}\gamma^2 K}{8} I_{T_{eff}}(T_m)
-  \frac{\pi g^2_{bs} \gamma^2 K}{4} T_{eff}I_{\eta}(T_m)
\label{rg3a}\\
\frac{dT_m}{d\ln{l}} = -T_m \label{rg4a}
\end{eqnarray}

where to summarize:
\begin{eqnarray}
I_{T_{eff}}(T_m\Lambda)= \int_{-2T_m\Lambda}^{2T_m\Lambda} d\bar{\tau}
{\rm Re}\left[
B(T_m\Lambda,\bar{\tau})\right]\\
I_{\eta}(T_m\Lambda)= -\int_{-2T_m\Lambda}^{2T_m\Lambda} d\bar{\tau}
\bar{\tau}{\rm Im}\left[
B(T_m\Lambda,\bar{\tau})\right]
\end{eqnarray}
where ${\rm Re}[B]=(B+B^*)/2$, ${\rm Im}[B]$=$(B-B^*)/(2i)$, and
\begin{eqnarray}
B(T_m\Lambda,\bar{\tau})=C_{+-}(0,T_m\Lambda,\bar{\tau})F(T_m\Lambda,\bar{\tau})
\end{eqnarray}
with $C_{+-}(0,T_m\Lambda,\bar{\tau})=\langle e^{i\gamma \phi_+(0,\bar{\tau}+T_m\Lambda/2)}e^{-i\gamma \phi_-(0,\bar{\tau}-T_m\Lambda/2)} \rangle$.
This quantity within leading order in perturbation theory is,
\begin{eqnarray}
&&C_{+-}(0,T_m\Lambda,\bar{\tau})
=\left[\frac{1}{\sqrt{1 +\bar{\tau}^2}}\right]^{2K_{neq}}
\left[\frac{\sqrt{1 + (2T_m\Lambda+\bar{\tau})^2}}{\sqrt{1 + (2T_m\Lambda)^2}}\frac{\sqrt{1
+ (2T_m\Lambda-\bar{\tau})^2}}{\sqrt{1 + (2T_m\Lambda)^2}}\right]^{K_{tr}}e^{-2i K_{eq}
\tan^{-1}\bar{\tau}}
\end{eqnarray}
while $F$ is given by
\begin{eqnarray}
&&F(T_m\Lambda,\bar{\tau}) = 2K_{neq}\left[\frac{1}{1 + \bar{\tau}^2}
\right] +2K_{tr}\left[\frac{1}{1 + \left(2T_m\Lambda\right)^2}
\right]- 2i K_{eq}\left[\frac{\bar{\tau}}{1 + \bar{\tau}^2}\right]
\end{eqnarray}

\subsection{Global Quench $g\neq0, K_0\neq K$ }

It is helpful to first look at $I_{T_{eff},\eta}$ at steady-state $T_m \rightarrow \infty$.
These may be evaluated analytically to give
\begin{eqnarray}
I_{T_{eff}}(\infty) = \left(\frac{4\pi}{2^{2K_{neq}}}\right)\frac{\Gamma\left(2K_{neq}\right)}
{\Gamma\left(K_{neq}+K_{eq}\right)\Gamma\left(K_{neq}-K_{eq}\right)}
\end{eqnarray}
whereas the expression for $I_{\eta}$ is
\begin{eqnarray}
I_{\eta}(\infty) =\left[ \frac{8\pi K_{eq}}{2^{2K_{neq}}\left(2K_{neq}-1\right)}\right]
\frac{\Gamma\left(2K_{neq}\right)}
{\Gamma\left(K_{neq}+K_{eq}\right)\Gamma\left(K_{neq}-K_{eq}\right)}
\end{eqnarray}
In the limit of small quenches, $I_{T_{eff},\eta}\rightarrow K_{neq}-K_{eq}\propto \left(K_0-K\right)^2$,
and therefore vanish when there is no bulk quench. This is consistent with the fact that for a LL which is
in its zero temperature ground state, the local non-linearity is incapable of producing any inelastic
scattering.

The generated dissipation and noise can be used to define an effective temperature~\cite{MitraGiamarchiPRL11,DallaTorreetalPRB2012}
\begin{eqnarray}
T_{eff}= \frac{g_{bs}^2 \gamma^2 I_{T_{eff}}}{4 \eta} \simeq g_{bs}^2\left(\frac{2\pi^2K_{eq}}{2^{2K_{neq}}}\right)\frac{\Gamma\left(2K_{neq}\right)}
{\Gamma\left(K_{neq}+K_{eq}\right)\Gamma\left(K_{neq}-K_{eq}\right)}
\end{eqnarray}
where we have approximated the local dissipation by its value in the absence of the non-linearity $\eta = 2/(\pi K)$.
The non-linearity gives corrections to $\eta$ which are small ${\cal O}(g_{bs}^2)$, however these
corrections diverge on approaching $K_{neq}=1/2$ as is apparent from the expression for $I_{\eta}(\infty)$.

The energy scale $T_{eff}$ is of the same order as the decay rate $\gamma_*$ of the OC correlator discussed in the
main text.
Note that the divergence of $I_{\eta}$ and hence of the local dissipation at $K_{neq}=1/2$ is also consistent with
the divergence of $\gamma_*$ observed in the decay of the OC correlator due to the back-scattering term.